%% file: ms.tex
\documentclass[apj]{emulateapj}
\usepackage{graphicx}

\slugcomment{Accepted by The Astrophysical Journal}

\shorttitle{Warm Molecular Hydrogen Tail}

\begin{document}

\title{A Warm Molecular Hydrogen Tail Due to Ram Pressure Stripping of a Cluster Galaxy}

\author{Suresh Sivanandam, Marcia J. Rieke, and George H. Rieke\altaffilmark{1}}

\altaffiltext{1}{Steward Observatory, University of Arizona, 933 North Cherry Ave, Tucson, AZ 85721; suresh@as.arizona.edu, mrieke@as.arizona.edu, grieke@as.arizona.edu.}

\begin{abstract}
We have discovered a remarkable warm ($130-160$ K) molecular hydrogen tail with a H$_2$ mass of approximately $4\times10^7$ M$_\odot$ extending 20 kpc from a cluster spiral galaxy, ESO 137-001, in Abell 3627. At least half of this gas is lost permanently to the intracluster medium, as the tail extends beyond the tidal radius of the galaxy. We also detect a hot ($400-550$ K) component in the tail that is approximately 1\% of the mass. The large H$_2$ line to IR continuum luminosity ratio in the tail indicates that star formation is not a major excitation source and that the gas is possibly shock-heated. This discovery confirms that the galaxy is currently undergoing ram-pressure stripping, as also indicated by its X-ray and H$\alpha$ tails found previously.  We estimate the galaxy is losing its warm H$_2$ gas at a rate of $\sim 2-3$ M$_\odot$ yr$^{-1}$. The true mass loss rate is likely higher if we account for cold molecular gas and atomic gas. We predict that the galaxy will lose most of its gas in a single pass through the core and place a strong upper limit on the ram-pressure timescale of 1 Gyr. We also study the star-forming properties of the galaxy and its tail. We identify most of the previously discovered external H$\alpha$ sources within the tail in our 8$\mu m$ data but not in our 3.6$\mu m$ data; IRS spectroscopy of the region containing these H$\alpha$ sources also reveals aromatic features typically associated with star formation. From the positions of these HII regions, it appears that star formation is not occurring throughout the molecular hydrogen tail but only immediately downstream of the galaxy. Some of these HII regions lie outside the tidal radius of the galaxy, indicating that ram-pressure stripping can be a source of intracluster stars. 
\end{abstract}


\section{Introduction}

The morphologies, gas content, and star-forming characteristics of galaxies vary drastically from the cluster to the field environment (see \cite{boselli06} and the references therein). There are a range of possibilities regarding the observed gas deficiency and reduced star formation of cluster spirals. One possible explanation is that gas-rich field galaxies are transformed as they fall into a cluster, during which process they are stripped of their gas and their star formation is halted. A variety of processes, such as ram-pressure stripping \citep{gunn72} and tidal interactions between galaxies \citep{moore96} and with the cluster potential \citep{henriksen96}, could be at play. However, a complete picture of the transformation of galaxies as they enter the cluster environment is still missing because we do not understand the degree to which cluster specific processes affect the evolution of the galaxies within them. Several studies suggest that the relative importance of ram-pressure stripping and tidal interactions depends on the mass of the galaxy in question \citep{cortese07,haines06,haines08}. Theoretical work shows that the transformation probably starts in group environments \citep{kawata08}, but gas loss also occurs from gas-only stripping (ram-pressure) events in the cluster environment itself \citep{tonnesen07}. The relative roles of these environments are ambiguous and would be clarified by an improved understanding of the effects of ram-pressure stripping. Truncated H$\alpha$ disks and HI tails on galaxies with undisturbed stellar disks provide evidence for this process \citep{koopmann04,chung07}. However, our understanding of the physical nature of the stripping from such systems will be improved by additional direct measures of the interaction with the intracluster medium.
\par
 {\em Spitzer} can offer a unique view by observing the interaction between the ICM and galactic interstellar medium (ISM) directly. \cite{appleton06} serendipitously discovered powerful molecular hydrogen emission emanating from the intragroup medium (IGM) of Stefan's Quintet. This emission is remarkable because the H$_2$ dominates strongly over other line emission \citep{cluver10} and the IR continuum is weak. The emission is thought to arise from a high velocity ($\sim 1000$ km s$^{-1}$) collision between an intruding galaxy and a tidal arm \citep{guillard09}. \cite{appleton06} place constraints on the temperature components of the excited gas and its column density by measuring the pure rotational H$_2$ transitions S(0) thru S(5). The observed H$_2$ lines are shock-excited because the spectrum is dominated purely by molecular hydrogen lines and the surface luminosity of the H$_2$ is ten times that from the X-ray emission associated with the shock-front. They show that there are at least two temperature components in the molecular gas: a warm and dense component ($\sim$ 200K) and a hot and sparse component ($\sim$ 700K). A more recent work by \cite{cluver10} that uses {\em Spitzer} data covers a significant portion of the Stefan's Quintet and confirms and expands on the conclusions of \cite{appleton06}. 
\par 
Observing these molecular hydrogen lines (if they exist) around galaxies falling into clusters would probe the gas-stripping processes and would provide quantitative estimates of the amount of gas being stripped. An excellent candidate for this test is ESO 137-001, a $0.2 L_*$ SBc galaxy in the Norma cluster (Abell 3627). ESO 137-001 is a remarkable galaxy with coaligned X-ray and H$\alpha$ tails that extend approximately 70 and 40 kpcs, respectively \citep{sun06,sun07,sun10}. Its projected distance from the center of the Norma cluster is only 0.28 Mpc. Its low radial velocity (77 $\mathrm{km\:s^{-1}}$) with respect to the cluster as a whole means most of its motion is in the plane of the sky. Moreover, deep H$\alpha$ imaging reveals bright extraplanar knots along the H$\alpha$ tail that are most likely very large star-forming regions \citep{sun07}, suggesting the stripped gas may be forming stars. Deep R$_C$-band imaging also shows low surface brightness streams that are aligned with the X-ray tail but have a larger width \citep{woudt08}. All evidence suggests this galaxy is currently undergoing ram-pressure stripping.
\par
In this paper we report the discovery with \emph{Spitzer} of a warm molecular hydrogen tail that trails ESO 137-001. Detailed \emph{Spitzer} observations of this galaxy and its tail expand on our understanding of ram pressure stripping and permit the estimation of the resulting mass flow. The paper is structured as follows. In $\S$2, we discuss the details of our observations and data reduction. In $\S$3, we present the results of our analyses. In $\S$4 we discuss the implications of our results. Finally in $\S$5, we list our conclusions. For computing distances, we adopt the concordance cosmological model ($\Omega_\Lambda = 0.73,$ $\Omega_m = 0.27,$ and $H_0 = 71$ km s$^{-1}$ Mpc$^{-1}$). We adopt redshifts provided by the NASA Extragalactic Database of 0.0157 and 0.01544 for the cluster and galaxy, respectively. This yields a luminosity distance of 67.1 Mpc to the cluster and an angular scale of $1\arcsec= 0.316$ kpc. All reported errors in this work are quoted at the 1$\sigma$ level.

\section{Observations and Data Reduction}

In this paper we detect and characterize warm H$_2$ emission and resolve the morphology of the stripped gas in the tail of ESO 137-001, and study the star-forming properties of the galaxy and the extraplanar HII regions. To this end, we carried out IRAC imaging and IRS spectral mapping observations. Our IRAC imaging allows us to resolve star-forming regions and study the stellar properties of the galaxy. IRS spectral maps provide detailed information (at a coarser spatial resolution) about the physical properties of the ISM such as the strength of the aromatic features, molecular hydrogen lines, and fine-structure lines. The data were taken as part of the \emph{Spitzer} GTO program 50213 (PI: G. Rieke), as discussed below.

\subsection{IRAC Imaging}
We observed ESO 137-001 on May 10, 2008 in all four (3.6, 4.5, 5.7, and 8$\mu m$) IRAC channels \citep{fazio04}. Frame times were kept short to prevent the saturation of the array from Milky Way foreground stars because the galaxy is located at a low galactic latitude ($b\sim-7^{\circ}$). We obtained a total on source integration time of 260 s in each of the channels. We constructed IRAC mosaics of the galaxy with MOPEX (version 18.1.5) using the Basic Calibrated Data (BCD) from the Spitzer pipeline (version S17.2) as input and following the Spitzer Science Center (SSC) IRAC reduction cookbook. Bad pixel rejection was achieved through dithering and median filtering.

\subsection{IRS Spectral Mapping}

\begin{figure}
\epsscale{1}
\plotone{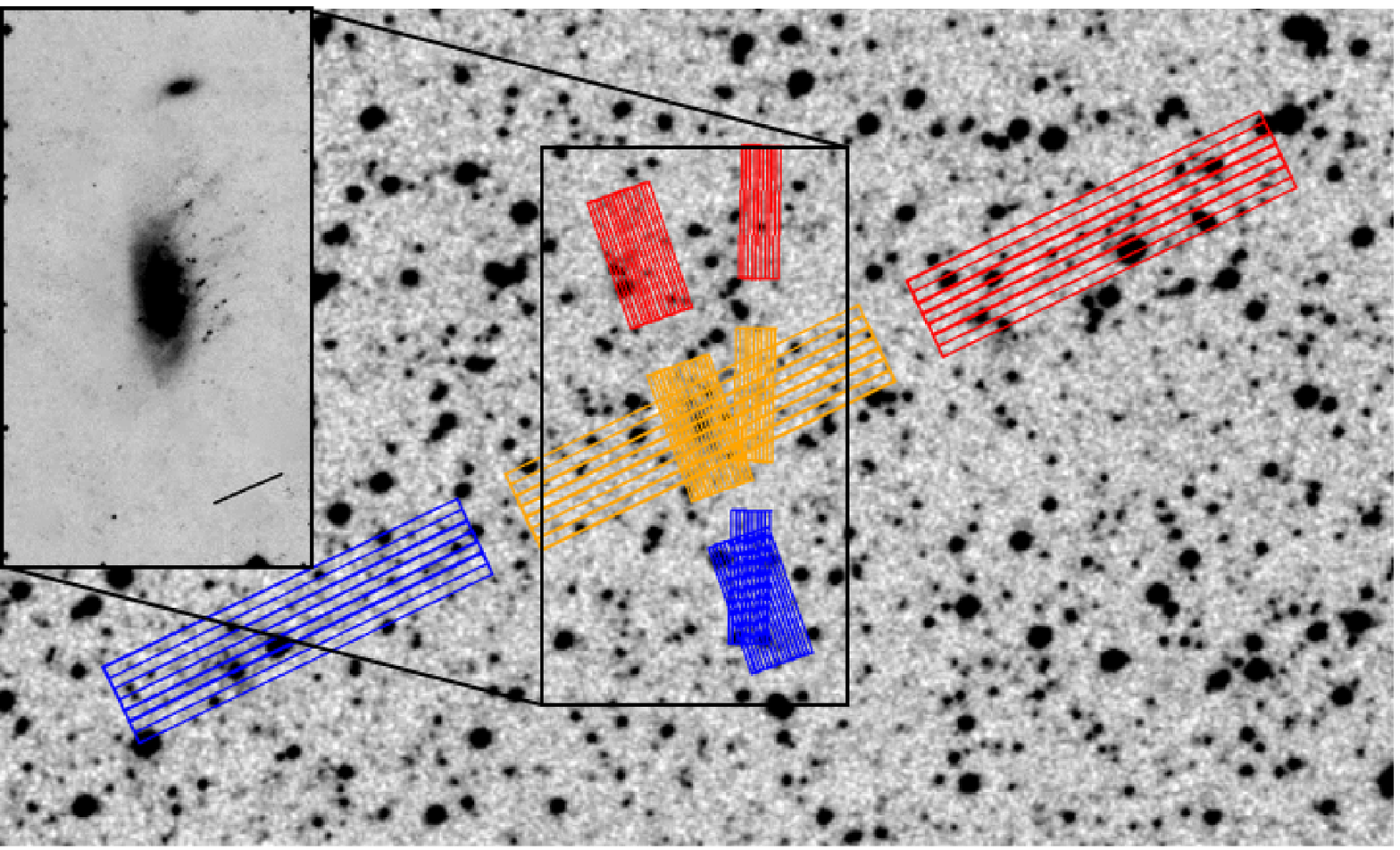}
\caption{6$\arcmin$ by 10$\arcmin$ Digital Sky Survey Image centered about ESO 137-001 with our SL/LL IRS pointings overlaid. North is up and East is left. SL slits are smaller than the LL slits. Regions with both 1st and 2nd order IRS pointings are shown in orange. 1st order only and 2nd order only pointings are shown in red and blue, respectively. Background subtraction is carried out by outrigger slit pointings. The inset is the 2.2$\arcmin$ by 4$\arcmin$ star-subtracted R$_C$-band image of ESO 137-001 reproduced from Figure 10 of \cite{woudt08}. This image has been contrast enhanced to reveal low surface brightness features. The diagonal line shown in the inset is the direction of elongation of the E/S0 population discussed in \cite{woudt08}, which lies very close in position angle to the X-ray and H$\alpha$ tails. We chose the long-low slits to be coaligned with the X-ray tail discovered by \cite{sun06}. \label{pointing}}
\end{figure}

The observations of ESO 137-001 used both the SL and LL IRS modules that together span 5.3 to 38.0 $\mu m$ \citep{houck04}. The full wavelength range allows us to observe several ground vibrational state H$_2$ rotational lines, specifically the $\nu$=0-0 S(0) thru S(7) transitions. We used the most recent pipeline reduced data (version S18.7) and \emph{CUBISM} (version 1.7) spectral map cube construction software \citep{smith07a} for all of our analyses to take advantage of the most up-to-data data calibrations. Our first set of IRS observations on May 2, 2008 was designed so each sky position was visited by two different IRS pixels --- the minimum required to produce a good spectral map. This was achieved by stepping the SL and LL slits in the perpendicular direction by 1.85$\arcsec$ and 5.1$\arcsec,$ respectively, such that the spectral map was centered about the galaxy. Moreover, during the time of our observations the LL slits were aligned to include the galaxy tail.  For our second set of IRS SL-only observations on April 6, 2009, we used the same mapping technique, but covered a region immediately west of the galaxy (presumably on the leeward side of the ICM wind). We used the outrigger off-source pointings from both LL and SL observations to measure the background.  Figure \ref{pointing} shows the final configuration of the IRS slit positions for both sets of observations. Final on-source integration times per pixel for our first set of IRS observations were 235 s, 235 s, 2265 s, and 880 s for the SL1, SL2, LL1, and LL2 spectroscopic cubes, respectively. The final per pixel integration time for our second set of IRS observations was 940 s for each of the SL1 and SL2 spectroscopic cubes. 
\par
Background subtraction was carried out using \emph{CUBISM}, which averages the frames of all off-source pointings associated with a given spectrograph using a min/max clip, and subtracts it from the on-source frames. If there is a bright source within one of the off-source pointings, the averaging technique of all off-source pointings diminishes the contribution of the source. The final background frame was visually inspected; the LL1, SL1, and SL2 (for both datasets) background frames showed no obvious signs of bright source contamination. However, the LL2 background revealed a few faint continuum sources that were on the south east side of the galaxy. Fortunately, these sources do not contaminate the regions from which we extract our spectra in this paper, and we do not attempt to mitigate this issue. All SL and LL spectrographs suffer from significant bad pixel issues, which we addressed through a manual process. We visually inspected the spectral cube for any artifacts along the direction of the slit steps. We manually masked the offending pixel that produced the artifact (identifying it using the backtracking feature in \emph{CUBISM}), rebuilt the spectral cube, and verified the artifact was removed. 
\par
The spectra from the different spectrograph orders were stitched together and placed on a single wavelength scale without any need for rescaling the flux of any individual order. Our processing of the data produces fluxes in the orders that are consistent with each other in the regions of overlap. As a check, we compared the IRS fluxes with an independent 8$\mu m$ IRAC flux measurement. For the spectral map centered around the galaxy, we found the 8$\mu m$ flux (derived from a weighted average using the IRAC 8$\mu m$ filter transmission as weights) of the final reduced spectra of a 22$\arcsec$ diameter aperture centered about the galaxy agreed with the IRAC 8$\mu m$ extended-source corrected flux to within 15\%. For the first IRS dataset, the sizes of the final spectral maps are $67\arcsec\times30\arcsec,$ $183\arcsec\times41\arcsec,$ and $20\arcsec\times28\arcsec$ for SL-only, LL-only, and both SL and LL maps, respectively. For the second IRS dataset, the sizes of the final spectral maps were $59\arcsec\times16.7\arcsec$, $39\arcsec\times18.4\arcsec$ (parallelogram shaped) for SL-only, and both SL and LL maps. When the LL data were combined with the SL data, we interpolated the LL data onto the SL data's pixel grid. \emph{CUBISM} generated error cubes associated with each spectral map. We used these error values in our spectral extractions unless otherwise noted. They do not include systematic errors, for example due to calibration errors or fringing.

\section{Results}

\subsection{Conditions in the Warm Molecular Hydrogen Tail}
\label{tailproperties}
\begin{figure}
\epsscale{0.8}
\plotone{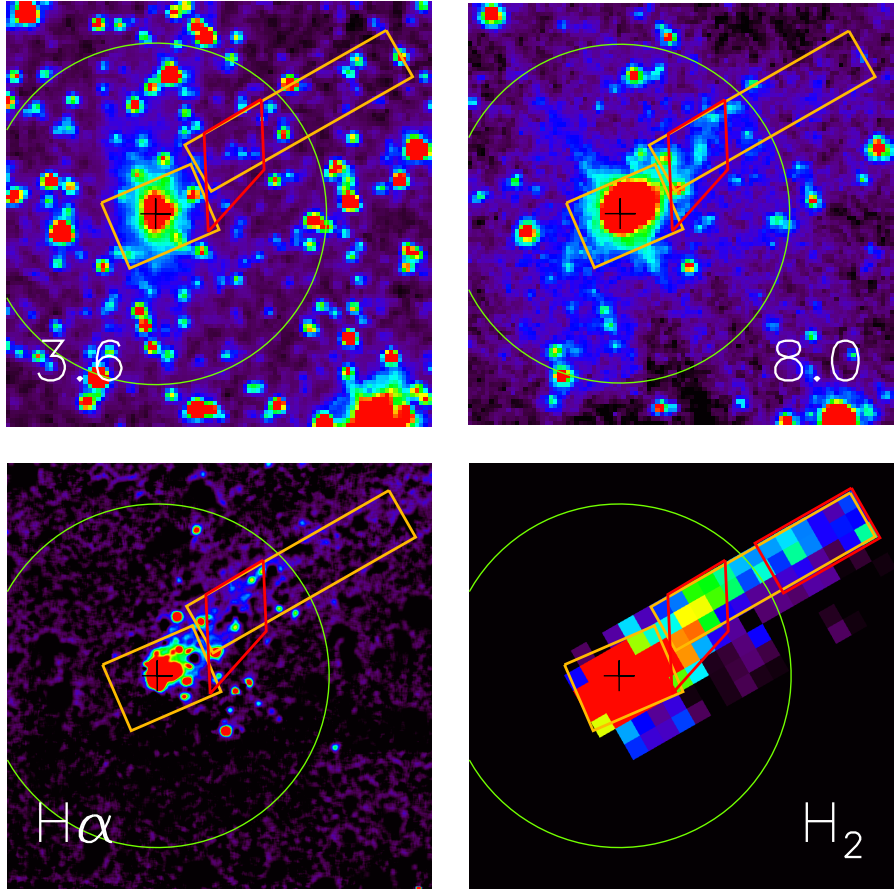}
\caption{2$\arcmin$ by 2$\arcmin$ images of ESO 137-001 at 3.6$\mu m$, 8$\mu m$, H$\alpha$ \citep{sun07}, and rest 17.035$\mu m$ (H$_2$ 0-0 S(1) transition) wavelengths. North is up and East is left. The H$\alpha$ image is smoothed to 1.2$\arcsec$ resolution. In the H$_2$ image, the emission shown is at least 3$\sigma$ above the background level. The colors follow the visible spectrum, where blue is represents the faintest emission, and red the brightest. The center of the galaxy is shown with a black cross. The polygons outline the spectral extraction regions. The box that is centered about the galaxy is termed the nuclear region and includes both short-low (SL) and long-low (LL) coverage. The red polygon includes full SL and LL coverage of the tail and is identified as the SL/LL tail region, while the long orange and smaller red rectangular boxes that enclose the molecular hydrogen tail have only LL coverage and are termed the LL full tail and far tail regions, respectively. The large green circle centered about the galaxy signifies the tidal radius. The 3.6$\mu m$ image shows a relatively undisturbed spiral galaxy. We observe a cometary structure at 8$\mu m$ that strongly suggests ram-pressure stripping. We also see a strong correspondence between the 8$\mu m$ and H$\alpha$ images, which is expected for star-forming regions. The morphologies at H$\alpha$ and 8$\mu m$ do not correspond to the 3.6$\mu m$ image, suggesting that the star-forming regions are relatively young.  A strikingly long molecular hydrogen tail is shown in the 17.035 $\mu m$ image. The position angles of the tails seen in H$\alpha$ and 8$\mu m$ are coaligned with the molecular hydrogen tail. This is evidence for warm gas being stripped from a galaxy with knots of star formation within it. \label{allbands}}
\end{figure}

\begin{figure}
\epsscale{0.8}
\plotone{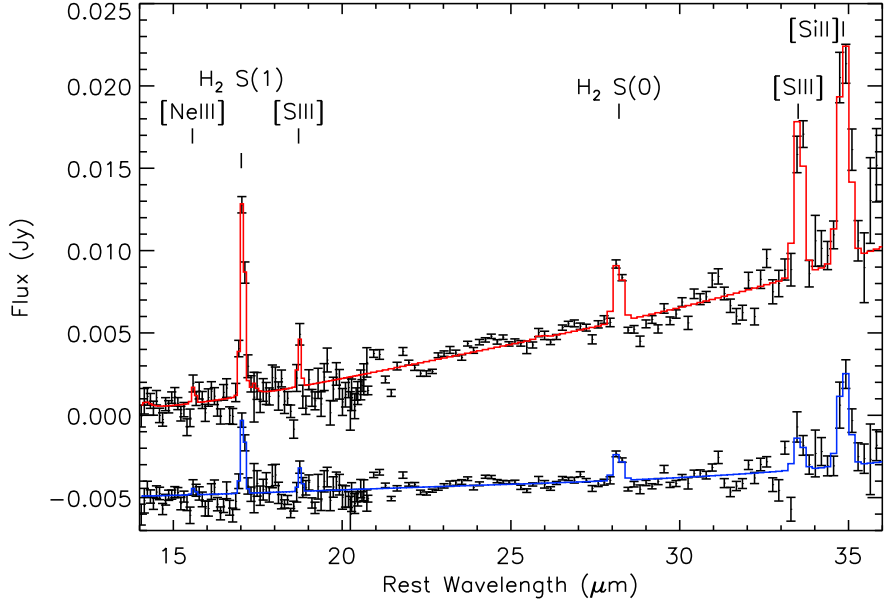}
\caption{ESO 137-001 LL-only full tail and far tail regions rest-frame spectra. The far tail region flux has been offset by -5 mJy for better visibility. The red and blue lines denote the best fit to the full tail and far tail data, respectively. The spectra show strong H$_2$ 0-0 S(1) and S(0) transitions in addition to atomic fine structure lines in both cases. The full tail spectrum reveals some continuum emission from warm dust, while continuum emission is virtually non-existent in the far tail case even though strong H$_2$ and fine structure lines are still present. This is no surprise because the full tail spectrum contains some of the extraplanar star-forming regions while the far tail region was chosen not to include any star-forming regions. \label{tailfit}}
\end{figure}

\begin{figure}
\epsscale{0.8}
\plotone{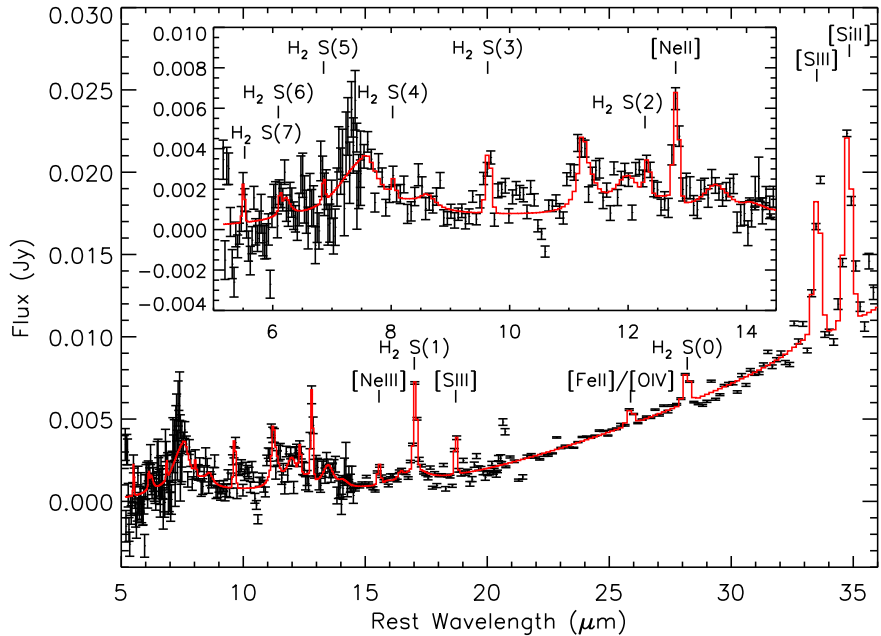}
\caption{ESO 137-001 SL/LL tail region rest-frame spectrum. The SL data are shown with a larger stretch within the inset plot. The red line denotes the best fit to the data. Atomic fine structure and H$_2$ 0-0 S(0) thru S(3), and S(7) lines are detected at greater than 3$\sigma$ significance within this region. The presence of aromatic emission indicates star formation is occurring in this region. \label{alltailfit}}
\end{figure}

In the lower right panel of Figure \ref{allbands}, we show the IRS image taken at the rest-frame 17.035$\mu m$ H$_2$ $0-0$ S(1) transition. We generated this image by summing up the flux from $17.2-17.4\mu m,$ and show only the emission that is 3$\sigma$ above the background. We find a remarkable tail emanating from ESO137-001 that extends all the way to the edge of the spectral map. This tail, aligned with both the H$\alpha$ and X-ray tails discussed by \cite{sun06,sun07}, is at least 20 kpc in length and approximately $3-5$ kpc in width. The virtually undisturbed stellar morphology of ESO 137-001 \citep{woudt08} plus the unusual state of the gas in the tail eliminate the possibility that it is a tidal feature (see Sections \ref{excitationmech} and \ref{physicalmech}). Instead, our observations support previous suggestions that the tail arises from ram-pressure stripping \citep{sun06,sun07}. We present spectra of this tail in Figures \ref{tailfit} and \ref{alltailfit}, which were generated by summing up the spectra within the LL-only (shown by the large orange and smaller red rectangles in Figure \ref{allbands}) and SL/LL (shown by the red polygon in Figure \ref{allbands}) regions, respectively. The large LL-only extraction aperture is called the LL-only full tail region, while the smaller LL-only aperture is called the LL-only far tail region. The solid angles ($\Omega$) of the SL/LL, LL-only full tail, and LL-only far tail are $9.07\times10^{-9},$ $2.38\times10^{-8},$ and $1.09\times10^{-8}$ sr$^{-1},$ respectively. The LL-only far tail region was chosen to be uncontaminated by star formation (see Section \ref{8umsec}) and outside the tidal radius of the galaxy. The spectra show strong detections of molecular hydrogen rotational (H$_2$ $\nu=0-0$) lines in these regions: the 5.511 $\mu m$ S(7), 9.665 $\mu m$ S(3), 12.279 $\mu m$ S(2), 17.035$\mu m$ H$_2$ $0-0$ S(1), and 28.22$\mu m$ H$_2$ $0-0$ S(0) transitions. Several fine structure lines and aromatic features were also observed within these apertures. 

\subsubsection{Molecular Hydrogen Tail Properties}
\begin{figure}
\epsscale{0.8}
\plotone{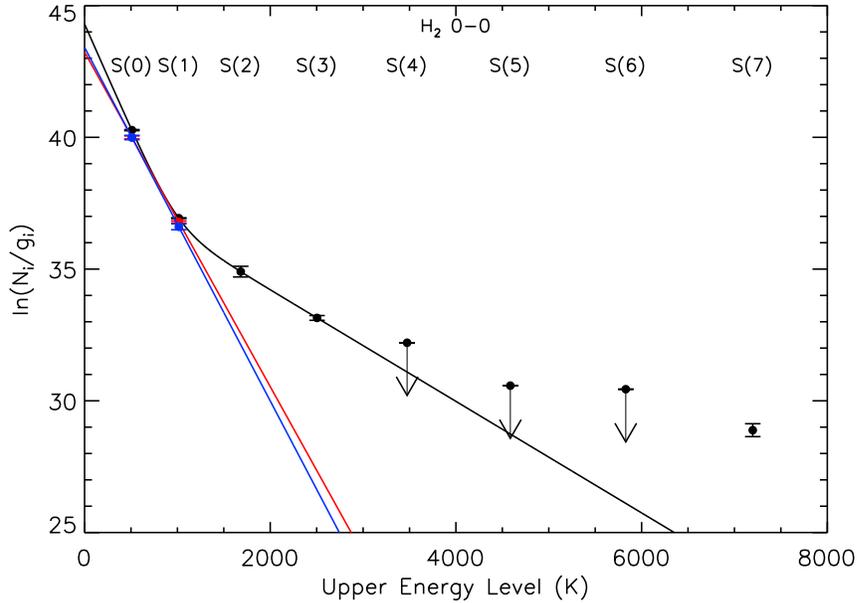}
\caption{Excitation diagram for the molecular hydrogen tail. The black, red, and blue points represent the populations in the SL/LL tail, LL-only full tail, and LL-only far tail regions, respectively. The LL-only full tail and far tail regions were fit by single temperature models, which are represented by the solid red and blue lines, respectively. The SL/LL region population was fit by a two-temperature (hot and warm components) model. For the two temperature model fits, the S(0), S(1), S(2), S(3) lines were included. The black curve is a two-temperature fit to the SL/LL tail data. The S(7) line is several standard deviations away from the best fit curve. A possible solution for this issue may be the inclusion of a third temperature component. However, we have insufficient data to constrain an additional temperature component. This is not a significant issue as most of the mass is contained with the lowest temperature component.
\label{excitation}}
\end{figure}

We calculate the physical properties of the stripped molecular hydrogen by assuming an optically-thin, single (or double) temperature gas model. We first compute the column density of each line, $N_{S(i)}$, using the following equation:
\begin{equation}
N_{S(i)} = \frac{\textrm{flux}_{S(i)}}{A_{S(i)}\times h \nu_{S(i)}} \times \frac{4\pi}{\Omega}
\end{equation} 
where flux is determined using \emph{PAHFIT} \citep[version 1.2;][]{smith07b}, which fits for both the continuum and lines, $A_{S(i)}$ is the Einstein A-coefficient for a given transition obtained from \cite{turner77}, $\nu_{S(i)}$ is the frequency of the transition (values are tabulated in \cite{black87}), and $\Omega$ is the solid angle of the aperture used. The errors for the line fluxes are obtained from 1$\sigma$ errors reported by the fitting algorithm. We do not include any extinction correction in the fit of the lines because it is likely to be low in the tail region outside of the galaxy. We present the excitation diagram ($\ln(N_{S(i)}/g_u)$ versus $T_{u}$) from the observed ground-state rotational transitions in the LL-only regions and SL/LL region in Figure \ref{excitation}. $g_u$ and $T_u$ represent the degeneracy and transition energy expressed in terms of the temperature of the upper level of the transition. The degeneracy g for molecular hydrogen can be expressed as $(2J+1)$ for even/para states (i.e. $J = \textrm{even}$) and $3(2J+1)$ for odd/ortho states (i.e. $J = \textrm{odd}$). We have firm detections of the S(0) and S(1) lines in the LL-only region and S(0), S(1), S(2), S(3), and S(7) lines in the SL/LL region. From the excitation diagram, we can infer that there are multiple temperature components in the molecular hydrogen tail within the SL/LL region.
\par
Due to the lack of electrical dipole moment in the hydrogen molecules, the only allowed radiative transitions between rotational states require $\Delta J = \pm 2$. Therefore, ortho and para states are radiatively decoupled. The ortho/para ratio (OPR) of molecular hydrogen in local thermodynamic equilibrium (LTE) varies from $\sim 1.5$ at 100 K to $\sim 3$ above $200 K$ \citep{burton92}. The lack of radiative coupling requires collisions between molecules (and/or a catalyst) to reach LTE. The warm component of the molecular hydrogen has temperatures that typically fall within this range of temperatures. Consequently, the OPR may not conform with the LTE value if the hydrogen molecules formed at temperatures higher than 200 K and have now cooled to temperatures lower than 200 K without sufficient density to reach LTE. The OPR is then frozen to the value when the gas was last in LTE. We carry out a test outlined by \cite{roussel07} to check if the observed population is not in LTE. They characterize any deviation from LTE with the $OPR_{\textrm{high}\:T}$ parameter, which is the OPR at high gas temperatures. At LTE, $OPR_{\textrm{high}\:T}$ is by definition 3. \cite{roussel07} state that if the following condition is satisfied the gas is consistent with a population in LTE: $T(S(0)-S(1)) \leq T(S(0)-S(2)) \leq T(S(1)-S(2)) \leq T(S(1)-S(3)) \leq T(S(2)-S(3))$ where $T(S(i)-S(j))$ is the temperature derived from $S(i)$ and $S(j)$ line ratios. The line ratios of lower excitation energy lines probe components at lower temperatures in a gas with a distribution of temperatures. Temperatures derived from $T(S(0)-S(2))$ and $T(S(1)-S(3))$ are independent of $OPR_{\textrm{high}\:T}$, whereas the rest depend on it. The temperatures derived from our lines satisfy the above equality and are consistent with being in LTE. This does not mean the population is definitively in LTE, as it is possible to have slightly lower values of the $OPR_{\textrm{high}\:T}$ ratio and still satisfy the condition. However, we assume LTE in our fits of the molecular hydrogen population.
\par
To quantify the physical properties of the gas, we first determined the temperature and total column density ($N_{tot}$) of the LL-only regions by fitting the S(0) and S(1) lines using Equations \ref{1model} and \ref{ntot}: 
\begin{equation}
\ln(N(T_u)/g_u) = \ln(N_{0})-\frac{T_u}{T}
\label{1model}
\end{equation}
where $N(T_u)$ is the column density of a transition, $g_u$ is the degeneracy of the transition, $N_0$ is the column density of the $J=0$ state, $T_u$ is the energy (in Kelvin) of the upper energy level of the transition, and T is the temperature of the gas.
\begin{equation}
N_{tot}(T) = \sum_{j=0}^{j=25} g_j N_0\exp\left(-\frac{T_{j}}{T}\right)
\label{ntot}
\end{equation}
where $N_{tot}$ is the total column density of the gas, $j$ is the rotational state, $T_{j}$ is the energy of that state. $T_0 = 0$ K, and $T_1 = 170.6$ K for molecular hydrogen.
Equation \ref{ntot} was summed up to the H$_2$ 0-0 $J=25$ state, which corresponds to an excitation energy of 37728 K, well above the observed temperatures. Given the limited information, we fitted a single temperature of $157\pm3$ K and a total column density of $2.46^{+0.25}_{-0.23}\times10^{19}$ cm$^{-2}$ in the LL-only full tail region. The errors in the temperature and column density were determined through Monte Carlo runs. 10,000 fitting trials were calculated where the trial line fluxes were drawn from a gaussian distribution with a mean and a full-width-half-max (FWHM) that were defined by the measured line flux and line flux error, respectively. We derive a total H$_2$ mass within the tail of $4.11^{+0.43}_{-0.39}\times10^7$ M$_\odot.$ In Figure \ref{excitation}, the blue line is the best fit single-temperature model. For the LL-only far tail region, we obtain a temperature of $149\pm6$K and a total column density of $2.90^{+0.51}_{-0.43}\times10^{19}$ cm$^{-2}.$ This corresponds to a total H$_2$ mass of $2.15^{+0.38}_{-0.32}\times10^7$ M$_\odot.$ 

\par
Next, we characterized the molecular hydrogen population in the SL/LL region. As shown by Figure \ref{excitation}, the additional information requires a model with multiple components with different temperatures. We chose the simplest model consisting of two temperature components as given in Equation \ref{2model}.
\begin{equation}
\begin{array}{l}
\ln(N(T_u)/g_u) =  \ln(N_{0,1})+ \\
\:\:\:\:\:\:\:\:\:\: \ln\left[\exp\left(-\frac{T_u}{T_1}\right)+f_{2,1}\exp\left(-\frac{T_u}{T_2}\right)\right]
\end{array}
\label{2model}
\end{equation}
where $N_{0,1}$ is the column density of the $J=0$ state of the first component, $T_1$ and $T_2$ are the temperatures of the first and second components, respectively, and $f_{2,1} = N_{0,2}/N_{0,1}$ where $N_{0,2}$ is the column density of the $J=0$ state of the second component. We only fit the S(0), S(1), S(2), and S(3) lines. We did not include the S(7) line in our fits for two reasons: 1.) we would require a third temperature component to account adequately for this line, but we do not have sufficient line detections to constrain a third component; and 2.) this line is most affected by UV fluorescence, which is a result of star formation. Our spectral maps showed some evidence for a patchy distribution of the S(7) line emission, though no clear correlation was observed with star-forming regions. The best fit model (shown in black in Figure \ref{excitation}) provides temperature estimates of $125^{+10}_{-19}$ K and $472^{+77}_{-56}$ K for the two components. The total column density for each component was calculated using Equation \ref{ntot}, producing a value of $5.9^{+4.2}_{-1.1}\times10^{19}$ cm$^{-2}$ for the warm and $5.9^{+4.4}_{-2.6}\times10^{17}$ cm$^{-2}$ for the hot components. These values correspond to masses of $3.6^{+2.6}_{-0.7}\times10^7$  M$_\odot$ and $3.6^{+2.7}_{-1.6}\times10^5$ M$_\odot$ for the warm and hot components, respectively. The results of all fits are tabulated in Table \ref{gasmass}. These results agree well with the hydrodynamical simulations of ram stripping by \cite{roediger08}. They predict that massive stripped tails can be produced even 0.5 to 1 Mpc from the cluster core (the projected distance of ESO 137-001 is 0.3 Mpc) with column densities of $\sim 10^{19}$ cm$^{-2}.$
\par
We also calculate the total H$_2$ line luminosity in each extraction aperture. To make this measurement, we must make a small extrapolation to include the luminosity from ground state rotational lines that are at shorter wavelengths than those already detected. We carry this out using the following equation:
\begin{equation}
L(H_2) = \Omega D_A^2 \left[ \sum^{i=23}_{i=0}N_{S(i)} A_{S(i)} h\nu_{S(i)}\right] \label{h2lumin}
\end{equation}
where $\Omega$ is the solid angle of the aperture, $D_A$ is the angular size distance to the galaxy, $N_{S(i)}$ is the column density of the molecules at the upper energy state for the $S(i)$ transition, and A and $\nu$ are the Einstein A-coefficient and frequency associated with that transition. The column densities are calculated from values obtained from the excitation diagram fit. For the SL/LL region, we obtain a luminosity of $1.59\times10^{40}$ ergs s$^{-1}$, while for the LL full tail and far tail regions, we obtain $1.36\times10^{40}$ and $0.59\times10^{40}$ ergs s$^{-1}$, respectively. Even though the LL full tail region covers the entire tail (i.e. larger solid angle than the SL/LL region), its H$_2$ luminosity is lower than the SL/LL region because of the significant emission coming from the hot component within the SL/LL region. If there is a hot component associated with either of the LL regions, then the stated luminosities are lower limits. Furthermore, there may be a third even hotter component inside the SL/LL region as revealed by the poor fit of the S(7) line, which suggests that even the luminosity derived for the SL/LL region is a lower limit. When compared to other wavelengths, we obtain luminosity ratios of $L(H_2)/L(H\alpha) \sim 1$ and $L(H_2)/L(\textrm{X-ray})\sim 0.1$ where the $L(H\alpha)$ and $L(\textrm{X-ray})$ are obtained from \cite{sun07} and \cite{sun06}, respectively. It is clear H$_2$ is a significant coolant. These ratios are only approximate, as the apertures over which the comparisons are made are not completely matched; the H$_2$ aperture is smaller than those used in the H$\alpha$ and X-ray measurements, so its cooling effect may be underestimated.

\subsubsection{Comparison of Molecular Hydrogen Gas Mass}
\par
We compare our results with the other mass estimates of the tail derived from observations in the H$\alpha$-band \citep{sun06}, and the X-ray \citep{sun07}. To make these comparisons, we compute the average gas surface density to mitigate issues arising from mismatched apertures. The gas surface density derived from our molecular hydrogen study ranges from $0.4-0.9\:\:\textrm{M}_\odot\:\:\textrm{pc}^{-2}$. Our measurement of molecular hydrogen mass is relatively robust because it is reasonably independent of the tail geometry and does not require a free-parameter $f,$ the filling factor. Nevertheless, there are systematic issues in our mass estimate: 1.) the IRS slit coverage does not cover the full axial extent of the tail, which causes us to miss some of the H$_2$ emission at the edges; and 2.) it is unclear what fraction of the molecular gas the warm H$_2$ emission represents. \cite{roussel07} in their survey of warm molecular hydrogen in SINGS galaxies find that for galaxies with CO detections (43 galaxies in total) warm molecular hydrogen ($T > 100$ K), on average, represents only $\sim10\%$ of the cold molecular hydrogen. There is a very broad range of warm to cold gas ratios in the SINGS sample ranging from $0.6 - 45\%$ with a scatter for the full sample of $\sigma \sim 10\%$. It is unclear if we can apply the fraction found in SINGS galaxies to explain the molecular hydrogen that is being stripped when the underlying formation and heating mechanism may be different. Preliminary CO measurements of the shocked region of Stefan's Quintet suggest a higher fraction ($\sim 50\%$) of warm to cold molecular gas \citep{guillard10}. However, Stefan's Quintet is an extreme environment with exceptionally strong molecular hydrogen emission, so the observed fraction may not be representative of conditions in the warm H$_2$ tail of ESO 137-001. 
\par
Both the H$\alpha$ and X-ray estimates rely on additional assumptions that we do not require. Both assume a cylindrical distribution for the gas to estimate an average electron density, which is then used to arrive at the mass of the radiating gas. Moreover, both estimates depend strongly on $f$, which is unknown. Taking these effects into consideration, we compute a surface density of hydrogen by dividing the mass estimate by the angular extent of the tail and obtain $\sim 2.5\:f^{\frac{1}{2}}\: \textrm{M}_{\odot}\:\textrm{pc}^{-2}$ for both the H$\alpha$ and X-ray tails. For the X-ray mass value, we use the tail mass from \cite{sun06}. This is approximately a factor of $3-6$ higher than our estimate if we assume $f=1,$ but agrees well with ours for the plausible value $f=0.03-0.1,$ assuming all three tails have similar masses.
\par
The exact physical mechanism that excites the stripped gas to produce emission at these different wavelengths is not entirely clear, making it difficult to predict the different states of the gas (see Section \ref{excitationmech} for a discussion). One needs to carry out a full census of the stripped gas through CO and HI observations to obtain a complete picture. Cold atomic gas measurements of the galaxy have been carried out, but were only able to place an upper limit on the HI mass of $10^9$ M$_\odot$ \citep{sun07}. If we assume the typical values found in the SINGS sample, cold H$_2$ may be a significant component of the stripped gas, and its mass is critical for calculating an accurate mass loss rate for the galaxy (see Section \ref{gasstripping}).  

\subsubsection{Fine Structure Line Emission in Tail}
We also fit several common fine structure lines using \emph{PAHFIT} in the LL-only and SL/LL-only regions to study possible excitation mechanisms for the observed molecular hydrogen tail. The lines are highlighted in Figures \ref{tailfit} and \ref{alltailfit}, and the line fluxes are tabulated in Table \ref{finestructure} for all three spectral extraction regions. We detect [NeII] ($12.8\mu m$), [NeIII] ($15.6 \mu m$), [SIII] ($18.7$ and $33.5\mu m$), and [SiII] ($34.8\mu m$).  We also observe an emission line in the SL/LL region at rest 26 $\mu m,$ which is likely a blend of the [OIV] $25.89\mu m$ and the [FeII] $25.99\mu m$ lines. It is possible to distentangle which of the two lines dominates the blend. The [OIV] line and the [FeII] line have vastly different excitation potentials, 56 and 7.9 eV, respectively. It is likely that this line detection is dominated by the [FeII] line as the low [NeIII]/[NeII] ratio of 0.2 suggests the gas is not highly excited. \cite{cluver10} draw similar conclusions where they too observe a $26\mu m$ line in Stefan's Quintet. Further progress can be made by comparing the [SIII] ($33.5\mu m$)/[SiII] ratio. \cite{dale06} carried out an IRS survey of a variety of different types of active galactic nuclei (AGN) and nuclear and extranuclear HII regions. They found a strong correlation between the type of object and the [SIII]/[SiII] flux ratios. HII dominated spectra typically have higher [SIII]/[SiII] ratios than AGNs and therefore this ratio can be a good discriminator of the type of energy source exciting the gas. We evaluate where each of the three spectral extraction regions fall in this correlation. The SL/LL region has a [SIII]/[SiII] ratio of 0.75, which is in a region that is largely dominated by HII regions. The same is true for the LL-only full tail, which has a ratio of 0.67. However, the LL-only far tail region yields a much lower ratio of 0.32, which is in an area that is dominated by AGNs. This is another indication that the gas in this region may be excited by a mechanism other than star formation, as discussed in Section \ref{excitationmech}.

\subsection{Star Formation in ESO 137-001 and its Tail}
The H$\alpha$ image clearly shows that the galaxy is actively forming stars in the nuclear region and to a lesser extent along the tail. Our infrared data offer us a complementary look at the star formation activity. We detect $8\mu m$ excess sources, which are clearly associated with the H$\alpha$ sources found by \cite{sun07}. We measure the SFR of the galaxy, focusing on the nuclear region where we see the majority of the star formation as indicated by H$\alpha$ imagery and we have full wavelength coverage across $5-36 \mu m.$ Finally, we address two questions: 1.) Is ESO 137-001 peculiar in any way in its star-forming properties? 2.) How does star formation progress in the molecular hydrogen tail? 

\subsubsection{8 $\mu m$ Dust Emission}
\label{8umsec}
We present the results of our IRAC imaging in Figure \ref{allbands}. The H$\alpha$ image of ESO137-001 was kindly provided by M. Sun and has been discussed in detail by \cite{sun07}. There are striking differences between the 3.6 $\mu m$ and 8 $\mu m$ images. The 3.6$\mu m$ data shows what appears to be a normal spiral galaxy with significant contamination from foreground Milky Way stars. The star-subtracted, contrast-enhanced, deep R$_C$-band image of \cite{woudt08}, reproduced in Figure \ref{pointing}, shows this galaxy more clearly and reveals only minor low surface brightness perturbations that extend in the southwest direction. However, the 8$\mu m$ image shows a completely different view. We see a cometary structure centered about the galaxy within which are several point-sources. Excess 8 $\mu m$ flux often arises from warm dust rich in aromatic hydrocarbons, associated with star formation \citep{calzetti07}. 
\par
\begin{figure}
\epsscale{0.7}
\plotone{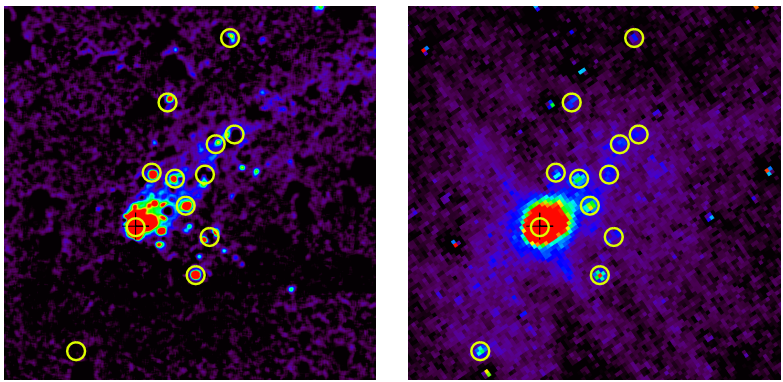}
\caption{2$\arcmin$ by 2$\arcmin$ H$\alpha$ image (left) of ESO 137-001 smoothed to 1.2$\arcsec$ resolution and the corresponding stellar continuum subtracted 8$\mu m$ image (right). The colors follow the visible spectrum, where blue is represents the faintest emission, and red the brightest. The black cross-hair indicates the center of the galaxy. Sources found by running a source detection algorithm on the 8$\mu m$ stellar continuum subtracted image are shown as yellow circles in both the H$\alpha$ and 8$\mu m$ image. The bright single pixel spots surrounded by a dark annulus are residuals from the stellar light subtraction. Every source present in the 8$\mu m$ image is present in the H$\alpha$ image with the exception of the point source southeast from the galaxy. It is unclear what this source is. The similarity between the two wavelengths establishes the H$\alpha$ and 8$\mu m$ sources are likely star-forming regions. These sources are in a fan-like configuration downstream of the galaxy. \label{Hacomp}}
\end{figure}

To understand the nature of the 8$\mu m$ sources, we first remove the 8$\mu m$ stellar flux to determine the emission from the aromatics. We achieve this by first convolving the 3.6$\mu m$ data to degrade the image quality to match the 8$\mu m$ resolution using kernels provided by \cite{gordon08}. Then we subtract the convolved 3.6$\mu m$ image after registering it to the 8$\mu m$ image and multiplying it by an arbitrary scaling ($\sim 0.3$) to ensure the bright stars are properly subtracted. The final subtracted image is presented in Figure \ref{Hacomp}. The kernel is imperfect because the wings of point-sources are over-subtracted; however, we are able to discern 8$\mu m$ excess clearly. We determine the position of the 8$\mu m$ excess sources using \emph{SExtractor} \citep{bertin96}. We first identify sources that have a minimum of 4 pixels detected more than 1.5$\sigma$ above the local background. We apply an additional 5$\sigma$ cut to these sources using the source flux and flux error generated by \emph{SExtractor} to generate the source catalogue. We visually inspect the final source catalogue and remove the sources that are associated with detector and stellar continuum subtraction artifacts. This method identifies all of the sources that can be identified by eye. We present the locations of the sources within 2$\arcmin$ of the galaxy in Table \ref{8umexcess} along with their distances from the center of the galaxy ($\alpha = $16:13:27.30 and $\delta =$ -60:45:50.5) obtained from the NASA Extragalactic Database.
\par
A comparison with the H$\alpha$ emission (see Figure \ref{Hacomp}) shows that the structures and point-sources seen at 8$\mu m$ are also present in the H$\alpha$ image. The 8 $\mu m$ excess source positions are overlaid on the H$\alpha$ image to indicate the correspondence. There are a few bright pixels surrounded by dark annuli seen at 8$\mu m$, which are in fact residuals from the stellar continuum subtraction. When we do a careful comparison of the two images almost every source at 8$\mu m$ within 1$\arcmin$ (approximately within the tidal radius) of the galaxy has a clear H$\alpha$ counterpart with one sole exception, a point source located southeast of the galaxy, also located on the side of the galaxy opposing the tail. It is unclear what this source is, and it is not very likely to be associated with the galaxy. The few fainter H$\alpha$ sources that are not seen in 8 $\mu m$ can attributed to the reduced sensitivity in the 8 $\mu m$ band due to its lower angular resolution. Overall, the source matching strongly suggests that these 8$\mu m$ excess sources are associated with the galaxy and are relatively young HII regions. This confirms \cite{sun07}'s hypothesis that these are star-forming regions (which they arrived at from their analyses of the B-band, I-band, and H$\alpha$ data), and possible sites for intracluster star formation.

\subsubsection{Measurement of Star Formation Rate in the Galaxy}
We calculate the star formation rate (SFR) in the central parts of the galaxy by first estimating the MIPS 24$\mu m$ \citep{rieke04} flux density from our IRS spectrum multiplied by the MIPS relative response curve. We find a MIPS 24$\mu m$ flux density of $45\pm2$ mJy in the nuclear extraction region, which contains almost all of the 24 $\mu m$ flux (see Figure \ref{24micron}). The uncertainty in the flux was determined by computing the standard deviation of the residual of a continuum fit from 20 to 30 $\mu m,$ excluding the S(0) line. This is necessary because the \emph{CUBISM} software underestimates the errors in that wavelength range and small ripples associated with fringing are present in the continuum. Hereafter, all MIPS 24$\mu m$ flux density measurements are carried out in this manner.
\par
 
Using \cite{rieke09}'s SFR-to-24$\mu m$ calibration produces a SFR value for the galaxy of 0.6 M$_\odot$ yr$^{-1}$. We note, however, our galaxy's 24$\mu m$ luminosity falls a factor of 2 lower than the valid range of calibration of the relation in \cite{rieke09}. By using their calibration, we likely underestimate the total (UV+IR) SFR somewhat because the UV leakage from the star-forming regions will be higher for lower 24 $\mu m$ luminosity galaxies. According to the Figure 7 of \cite{buat07}, the UV leakage can be about 15\% for our galaxy. After taking this into account, we deduce the total SFR (UV+IR) of ESO 137-001 to be 0.7 M$_\odot$ yr$^{-1}$. This value is dominated by the systematic uncertainty of the SFR-to-24$\mu m$ calibration $\sim 0.2$ dex \citep{rieke09}.
\par
Our SFR value is a factor of 2 lower than that measured by \cite{sun07}. \cite{sun07} used an extinction-corrected H$\alpha$ measurement to estimate a SFR of 2 M$_\odot$ yr$^{-1}$, which translates to 1.3 M$_\odot$ yr$^{-1}$ for the Kroupa IMF used by \cite{rieke09}.  There is a mismatch in the aperture used by \cite{sun07} and our work for the measurement of the SFR. We correct for this using the H$\alpha$ data provided by M. Sun to measure the relative difference of total H$\alpha$ flux between the two apertures. The difference is very small; the H$\alpha$ SFR measurement is reduced by 5\% when the 24$\mu m$ aperture is used yielding a H$\alpha$ SFR of 1.2 M$_\odot$ yr$^{-1}$. No statistical uncertainty for this measurement was quoted by \cite{sun07}; however, this measurement carries a systematic error of $\sim 0.15$ dex \citep{kennicutt98}. \cite{sun07} applied a 1.4 mag extinction correction estimated from the H$\gamma$/H$\beta$ line ratio. Reconciling the H$\alpha$ with the IR-determined SFR, however, would require that there is 0.8 mag of extinction at H$\alpha$. To probe this inconsistency, we use the Kroupa IMF SFR calibration given by \cite{kennicutt09} that uses both H$\alpha$ (not extinction corrected) and 24$\mu m$ luminosity. \cite{sun07} provide an extinction corrected L(H$\alpha$) within the region they calculate the SFR. After removing the extinction correction and applying the aperture correction, we obtain a L(H$\alpha$) of $6.5\times10^{40}$ ergs s$^{-1}$. Using this value along with the 24$\mu m$ luminosity yields a SFR of 0.7 M$_\odot$ yr$^{-1},$ which is consistent with our original 24$\mu m$ SFR measurement. This suggests that the extinction correction used by \cite{sun07} was probably overestimated.
\par
We now compare the SFR with the stellar mass of ESO 137-001 to see if it has an abnormal SFR. We first estimate the galaxy's stellar mass from its K-band magnitude. We obtained a K-band apparent magnitude of 11.6 from a catalogue of galaxies in Abell 3627 \citep{skelton09}, which corresponds to a $M_K = -22.5.$ This yields a stellar mass of $\sim 1.4\times10^{10}$ M$_\odot$ \citep{bell00}. A typical SFR for a cluster member of this mass is $\sim 5$ M$_\odot$ yr$^{-1},$ with a board distribution extending down to the rate in ESO 137-001 \citep{vulcani10}. Thus, there are no obvious anomalies in the star-forming properties of the galaxy, although it is less active than is typical.

\par
\begin{figure}
\epsscale{0.8}
\plotone{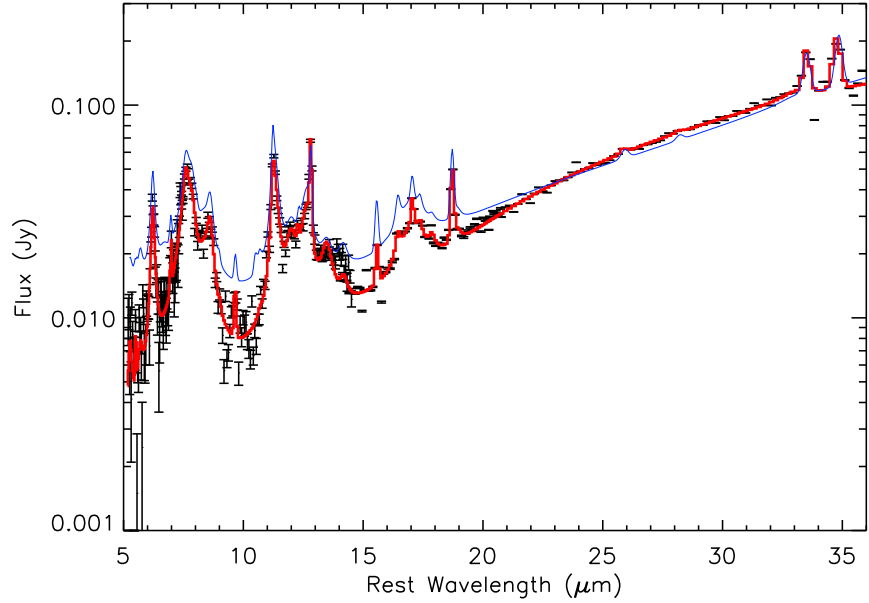}
\caption{ESO 137-001 nuclear region rest-frame spectrum. The red line denotes the best fit to the data. We have full short-low and long-low spectrograph coverage in this region. The blue line is a starforming galaxy template spectrum from \cite{smith07b}. The template spectrum was rescaled to match the 24 $\mu m$ flux of ESO 137-001. We see a fairly good match between the two spectra, suggesting there is nothing peculiar about the infrared properties of ESO 137-001. \label{nucfit} \label{template}}
\end{figure}

To explore the galaxy's star formation properties further, we look to its IR spectrum (see Figure \ref{nucfit}).  The galaxy unsurprisingly shows strong aromatic features and an infrared continuum, both clear indicators of on-going star formation. We compare the spectrum with templates provided by \cite{smith07b} and find it is well-matched with a star-forming galaxy template as shown by Figure \ref{template}. We also checked to see if there are any peculiarities in the aromatic features of ESO 137-001. We fit the nuclear spectrum of the galaxy using PAHFIT  \citep[version 1.2;][]{smith07b} with extinction correction, and the best fit is shown in Figure \ref{nucfit}. We tabulate the strength of various aromatic features in Table \ref{aromatic}. The derived flux ratios L($6.2 \mu m$)/L($7.7 \mu m$ Complex) of 0.3 and L($11.3 \mu m$ Complex)/L($17 \mu m$ Complex) of 2.1 are consistent with star-forming galaxies with HII-dominated nuclei \citep{smith07b}.

\subsubsection{Star Formation in the Molecular Hydrogen Tail}
\begin{figure}
\epsscale{1}
\plotone{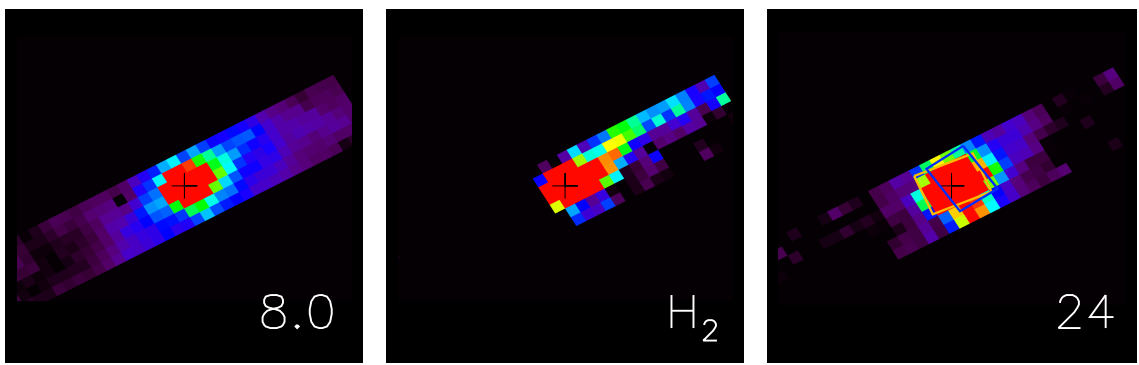}
\caption{2.5$\arcmin$ by 2.5$\arcmin$ images of ESO 137-001 at 8$\mu m$ (left), rest 17.035$\mu m$ (H$_2$ 0-0 S(1)) wavelength (center), and 24 $\mu m$. The colors follow the visible spectrum, where blue is represents the faintest emission, and red the brightest. In the 24$\mu m$ panel, the orange box is the spectral extraction region used to measure the SFR while the blue box is the region used by \cite{sun07} to compute their H$\alpha$ SFR. The 8$\mu m$ data was smoothed to match the 17$\mu m$ resolution after subtracting the stellar continuum using 3.6$\mu m$ data. The center of the galaxy is shown with a black cross. While there is a strong correspondence between the 8$\mu m$ and 24$\mu m$ emission, the molecular hydrogen tail is much more extended. This hints that the stripped molecular hydrogen must only have sufficient density to form stars shortly after it has been removed from the galaxy and that star formation cannot be the main source of excitation. \label{comp} \label{24micron}}
\end{figure}

The H$\alpha$ image \citep{sun10} demonstrates that star formation is occurring outside the galaxy near the head of the tail. To explore the star formation in this region, we compare the IRAC 8$\mu m$ data and the IRS 24$\mu m$ map with the molecular hydrogen S(1) line map. We first convolve the 8$\mu m$ continuum-subtracted image with a kernel we created using the 17$\mu m$ IRS PSF generated by stinytim and the 8$\mu m$ PSF using the same technique outlined above to degrade the H$\alpha$ image. In Figure \ref{comp}, we show that the S(1) emission is much more extended than the 8$\mu m$ emission, indicating that star formation only proceeds in the part of the tail immediately downstream of the galaxy. This is corroborated by our 24$\mu m$ map shown in Figure \ref{24micron}. We measure a MIPS 24$\mu m$-band flux of $3.5\pm0.5$ mJy within the SL/LL region where the majority of the extraplanar SF is occurring. The flux in this region is less than $10\%$ of the value at the galactic nucleus indicating that most of the SF is confined to the galaxy. Even though we detect 24$\mu m$ emission in the SL/LL region we do not quote a SFR. It is unclear how to translate the measured flux in the star formation rate as the ICM environment is hostile to dust, and the usual SFR calibrations may not hold.  We also confirmed the lack of MIPS 24$\mu m$-band flux detection ($< 1.10$ mJy, 3$\sigma$ detection limit) in the LL-only far tail region, where there is no associated 8$\mu m$ excess emission and half of the detected warm H$_2$ lies. This is also corroborated by the low [SIII]/[SiII] ratio in this region. We conclude a significant portion of the stripped molecular gas may not have high enough density to support star formation.  
\par
We also looked for any peculiarities in the aromatic features measured in the SL/LL region. We tabulate the fluxes in Table \ref{aromatic}.  The derived flux ratio f($6.2 \mu m$)/f($7.7 \mu m$ Complex) of 0.11 is peculiar for a star formation dominated region \citep{smith07b}. Its value is lower than the lowest value found by \cite{smith07b} of 0.2 for HII-dominated nuclei, which have a typical value of $\sim0.3$. However, the flux ratio f($11.3 \mu m$ Complex)/f($17 \mu m$ Complex) of 1.9 is not inconsistent with the average value observed of HII-dominated nuclei. One possible explanation for the observed discrepancy is that aromatic emission at $6.2 \mu m$ is somehow suppressed.

\section{Discussion}

\subsection{Excitation Mechanisms for Molecular Hydrogen}
\label{excitationmech}
\par
There are a number of mechanisms to excite molecular hydrogen: 1.) UV fluorescence \citep{black87}; 2.) X-ray irradiation arising when X-ray photons penetrate the molecular gas clouds deeply \citep{maloney96}; and 3.) shock-heating from high-velocity collisions. In the last case, a relatively fast J-shock can dissociate H$_2$ at the shock front only to have it form later on grain surfaces downstream. The lowest J rotational lines are collisionally excited in this process (see \cite{hollenbach89}). Recently, \cite{guillard09} have presented a sophisticated model of the H$_2$ emission from Stephan's Quintet. They are able to produce a multiphase medium that emits strongly in rotational molecular hydrogen lines, and to a lesser extent in X-ray and H$\alpha,$ from a large galactic scale shock caused by the collision of an intruding galaxy with a tidal arm. The shock-heated gas eventually cools to form molecular H$_2$, which is the main coolant in the gas, and is excited by low velocity, non-dissociative magnetohydrodynamic (MHD) shocks that dissipate most of mechanical energy of the collision. Their model adequately explains the two temperature molecular hydrogen component and the high $L(H_2)/L(IR)$ ratio discovered by \cite{appleton06}. This model qualitatively matches our observations, and we conclude that the H$_2$ tail can be shock-excited. 
\par
\begin{figure}
\epsscale{0.8}
\plotone{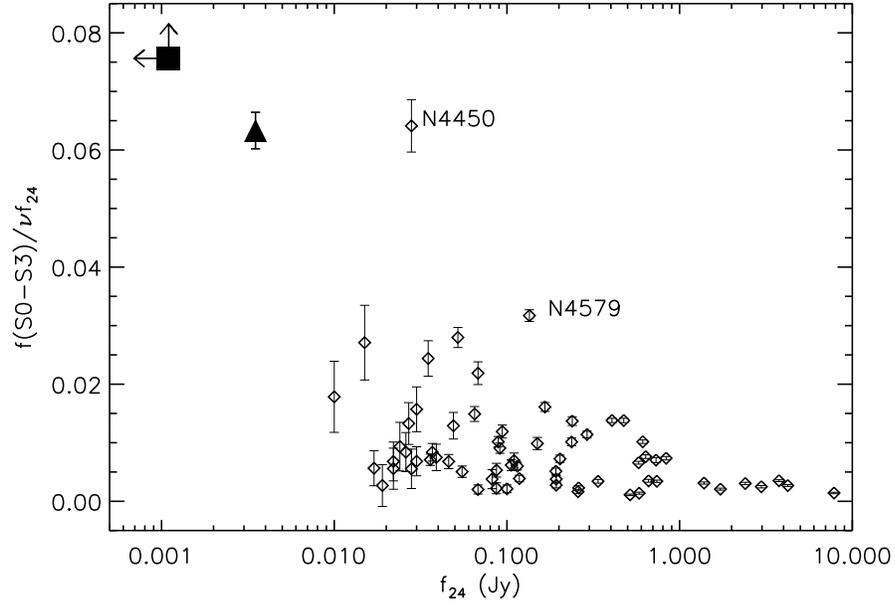}
\caption{A comparison of molecular hydrogen ground state rotational line flux (measured from S(0) thru S(3) transitions) to 24$\mu m$ flux as a function of 24$\mu m$ flux for SINGS galaxies \citep{roussel07} and warm molecular hydrogen tail. The diamond points represent the 66 galaxies from the SINGS sample. Two of the significant outliers in the sample, NGC 4450 and NGC 4579, are labelled in the plot. The warm H$_2$ in both NGC 4550 and NGC 4579 is thought to be heated through shocks from cloud collisions and/or X-ray irradiation because star formation and supernova remnants are unable to explain the high H$_2$ emission. The filled triangular and square points represent the values measured in the SL/LL and LL far tail regions of the warm molecular hydrogen tail, respectively. It is clear that the two measurements in the tail are significant outliers in this comparison sample and are comparable in value to the ratio for NGC 4450, which is likely heated by shocks through cloud collisions in addition to X-ray irradiation \citep{roussel07}.
\label{singscomp}}
\end{figure}

To investigate further the H$_2$ excitation mechanism, we compare our results with the warm molecular hydrogen inventory of the SINGS star-forming galaxies \citep{roussel07}. We use the molecular hydrogen flux to 24$\mu m$ flux ratio as a proxy for $L(H_2)/L(IR)$. The ratios of molecular hydrogen line flux (measured for the 0-0 S(0) thru S(3) lines) to MIPS $24\mu m$ flux for SINGS star-forming galaxies are shown as diamond points in Figure \ref{singscomp}. The errors in each data point shown are the errors in the molecular hydrogen line flux. Because they should be much smaller, the 24$\mu m$ flux errors were not quoted and therefore they are not included in the error in the ratio. We also calculated the same ratios for the SL/LL and the LL far tail regions, which are shown by the triangular and square points respectively. The S(2) and S(3) line fluxes for the LL region were calculated by extrapolating the temperature fit using Equation \ref{h2lumin}. The LL region derived ratio is a lower limit, as we only have an upper limit for the 24$\mu m$ flux. The ratios calculated for the tail regions are clearly significant outliers when compared to the SINGS sample even in the case of the SL/LL region of ESO 137-001 where there is contamination from star formation. In fact, the ratios are comparable to or greater than  the most significant outlier in the SINGS sample: NGC 4450.  \cite{roussel07} state that their two most significant outliers, NGC 4550 and NGC 4579, have significant excess in H$_2$ emission that cannot be explained by emission from PDRs or supernova remnant shocks. They state that shocks produced by cloud collisions in combination with X-ray irradiation could explain the energetics of the observed H$_2$ emission.
\par
We summarize several pieces of evidence that suggest that some process other than star formation must be the source of the H$_2$ excitation: 1.) the most compelling evidence is that approximately 50\% of the warm molecular hydrogen gas mass, which resides outside the tidal radius of the galaxy and within the hot ICM, has no detected star formation activity; 2.) even the region within the tail with some star formation activity shows unusually high luminosity of H$_2$ for its IR luminosity in comparison with the average value of SINGS galaxies; 3.) the H$_2$ line emission is at least 10\% of the observed X-ray luminosity and approximately similar to the H$\alpha$ luminosity of the tail, which suggests a highly energetic process heating the gas; and 4.) the [SIII]/[SiIII] ratio in the tail outside the tidal radius is consistent with gas excited by an AGN, possibly suggesting that the energy source may be hard. We conclude that the H$_2$ tail is likely stripped gas from the galaxy that is either heated by X-rays and/or through shocks produced by the galaxy's interaction with the ICM. We may have directly detected the molecular gas from on-going ram pressure stripping of an infalling cluster spiral. This possibility is discussed further in Section \ref{gasstripping}.

\subsection{Physical Mechanism Responsible for the Tail}
\label{physicalmech}
ESO 137-001 has a relatively normal spiral-like stellar distribution and asymmetric gas distribution as evidenced by the X-ray and H$\alpha$ tails. The existence of the molecular hydrogen outside the plane of the galaxy and its anomalous excitation mechanism raise several interesting questions: How did this molecular gas find its way out of the galaxy? What does this say about galaxy transformation in the cluster environment? How long does this phenomenon persist? To answer these questions we look at the possible processes that strip galaxies of their gas in clusters. We discuss why tidal stripping processes are not significant, and why ICM gas-stripping is the most likely scenario that explains the observed phenomenon.

\subsubsection{Galaxy-Galaxy Interactions}
Galaxy-galaxy interactions can be strong transformative processes in the cluster environment. These interactions can be classified into two broad categories: low speed and high speed. Low speed interactions frequently lead to significant morphological changes and mergers, whereas high speed encounters lead to tidal stripping of gas and stars from the outer edges of disks and induce nuclear starbursts. 
\par
If virialized, the high velocity dispersion of Abell 3627 \citep[$\sigma = 925\: \textrm{km s}^{-1}$;][]{woudt08} makes it an unlikely site for a low velocity interaction between two galaxies. However, the case of Abell 3627 is not simple. X-ray imaging of the cluster reveals a subgroup near its center, which is identified in X-ray residuals after subtracting a radially symmetric component. The cluster's diffuse ICM emission is elongated with a position angle of approximately $130^\circ$ \citep{boehringer96}. The central cD galaxy that lies within the subgroup has a peculiar velocity of 561 $\textrm{km s}^{-1}$ with respect to the cluster mean velocity indicating that the cluster is undergoing a merger at its center \citep{woudt08}. However, ESO 137-001 lies outside of the subgroup. There is also no compelling evidence that ESO 137-001 is part of a dynamically cooler subgroup \citep{woudt08} making it unlikely for it to have experienced a low velocity encounter with a nearby galaxy. Furthermore, the closest galaxy located within 10$\arcmin$ with the smallest difference in radial velocity with ESO 137-001 is ESO 136-024, which has a $\Delta v_r = 120 \:\textrm{km s}^{-1}$. One would expect tidally disrupted material to follow the trajectory of the perturber, but the location of ESO 136-024 is not aligned with the tail. In addition, there is no evidence in the stellar morphology \citep[][see Figure \ref{pointing}]{woudt08} for distortions in the direction of ESO 136-024.
\par
Galaxy harassment in the form of multiple fast interactions in clusters can strip gas and stars from small spiral galaxies  and turn them into dwarf elliptical or dwarf spheroidal galaxies \citep{moore96}. Simulations show that for high surface brightness spiral galaxies with stellar disk scale lengths of 3 kpc, a value comparable to that measured by \cite{sun07} for ESO 137-001, the effect of high speed encounters with massive galaxies is likely to be minor \citep{moore99}. In their simulations of harassing tidal interactions involving high surface brightness model galaxies similar to ESO 137-001, \cite{moore99} showed that the disk lost only $1-2$\% of its stars and largely retained its disk shape while its disk scale height increased by a factor of $2-4$ and its disk scale length decreased somewhat. The current disk morphology of ESO 137-001 suggests that if it had an interaction with a massive galaxy, it is likely to have had a minor impact on its stellar distribution. There is no physical reason why the gas would be preferentially stripped over stars in a tidal interaction. Moreover, other simulations show that in tidal interactions, gas typically tracks stars in tidal tails whereas ram-pressure can produce a significant difference between the stellar and gas distribution \citep{vollmer03, mayer06}. In fact, simulations show that tidal interactions often lead to gas inflow and bar formation, and ram-pressure is necessary to strip the gas \citep{mayer06}. Additionally, there is only a single tail, whereas both leading and trailing tails are usually present in a tidal interaction \citep{moore99}. However,  simulations show that it may be possible for ram pressure to erase tidal features produced by a single close encounter and produce long gaseous tails trailing the galaxy \citep{kapferer08}. Therefore, the clear difference between the distribution of stars within ESO 137-001 and the gas that trails behind it suggests that tidal interaction between galaxies is not the main contributing factor for the existence of the tail. 

\subsubsection{Galaxy-Cluster Tidal Interactions}
\label{clustertidalsec}
Inside the cluster core, the deep cluster gravitational potential can also foster galaxy transformation \citep{henriksen96}. When the cluster tidal force exceeds the galaxy's own restoring force, gas and stars can be tidally stripped from the galaxy and trigger a nuclear starburst \citep{henriksen96, cortese07}. We calculate the effect of the cluster potential for our galaxy by using the formalism outlined by \cite{henriksen96}. The cluster tidal effect can be expressed as a radial and a transverse acceleration, $a_r$ and $a_t$, respectively, which are given below:
\begin{equation}
a_t(r) = GM_{cl}(r)\frac{R}{\left[R^2+(r+R)^2\right]^{3/2}}
\end{equation} 
and
\begin{equation}
a_r(r) = GM_{cl}(r)\left[\frac{1}{r^2}-\frac{1}{(r+R)^2}\right]
\end{equation}
where $M_{cl}$ is the mass of the cluster, $r$ is the distance of the galaxy from the cluster center, and R is the radius of the galaxy. We can calculate the significance of the cluster tidal force by comparing it to the centripetal acceleration of the galaxy, $a_{gal},$ at the outermost parts of its disk:
\begin{equation}
a_{gal}=\frac{GM_{gal}}{R^2}
\end{equation}
where $M_{gal}$ is the dynamical mass of the galaxy, and $R$ is the physical extent of the galaxy. On one hand, if the radial tidal acceleration exceeds the centripetal acceleration of the galaxy, we would expect the gas and stars to be stripped by the cluster's gravitation force. On the other hand, if the transverse tidal acceleration exceeds the centripetal acceleration, the gas clouds within the galaxy will be compressed leading to a burst of star formation. 
\par
To determine the strength of the cluster tidal acceleration, we first estimate the cluster dynamical mass profile. We tested two approaches to determine the mass: one that uses the X-ray derived dynamical mass profile \citep{boehringer96}, and the other that extrapolates to smaller radii using an NFW profile \citep{navarro96} and the dynamical mass estimate determined by \cite{woudt08} at large radii. To find the best-fit NFW profile, we use the dynamical mass values for the cluster at 3 different radii (0.67, 1.35, and 2.02 $h^{-1}_{73}$ Mpc) from \cite{woudt08}. We use only the virial theorem ($M_{VT}$) and robust virial $M_{RVT}$ dynamical mass estimates at these radii for our fit. \cite{woudt08} suggest that their third method for mass estimation, projected mass estimator, is affected most by substructure, which is present in Abell 3627, and produces a mass estimate that is 50\% larger than $M_{VT}.$ $M_{VT}$ and $M_{RVT}$ are least affected by substructure and agree to within 5\%, which is why we fit to the average of these two values. Nevertheless, these masses are also likely to be overestimates due to the on-going merger in the cluster. The NFW parameters we obtained were $r_s = 346$ kpc and $c = 6.2.$ We do not quote errors on these parameters because error estimates were not given for the mass estimates.
\par
\begin{figure}
\epsscale{0.9}
\plotone{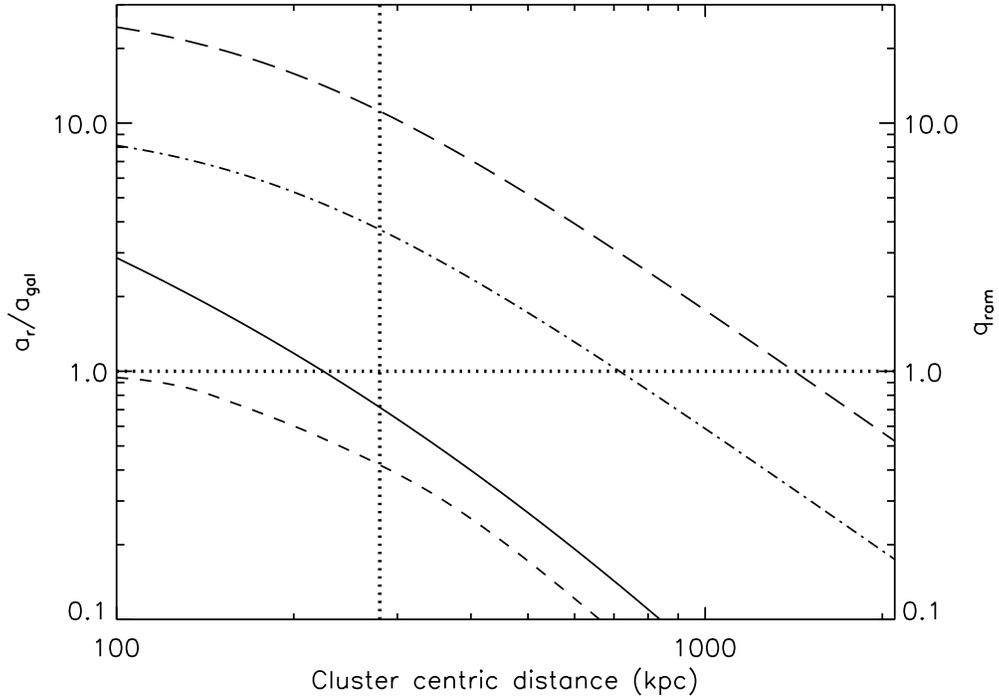}
\caption{Ratio of cluster radial $a_{r}$ and galaxy centripetal $a_{gal}$ acceleration ($a_r/a_{gal}$) and effectiveness of ram-pressure ($q_{ram}$) as a function of cluster-centric distance. The solid and short dashed curves represent $a_r/a_{gal}$ determined from a best-fit NFW cluster mass profile and a X-ray-data-derived mass profile, respectively. If the ratio is greater than 1 (the dotted horizontal line), the cluster tidal force will be sufficient to tidally strip the galaxy. However, at the projected distance of 280 kpc (the dotted vertical line), the minimum distance the cluster galaxy can be from the center, the $a_r/a_{gal}$ is less than 1 for both cluster mass profiles, indicating that tidal stripping from the cluster gravitational field is not a significant effect. The dash dotted and long dashed curves represent $q_{ram}$ for a galaxy traveling at 925 and 1602 km s$^{-1},$ respectively. When $q_{ram} > 1$, the \cite{gunn72} criterion is satisfied and ram-pressure stripping becomes effective. It is clear that ram-pressure stripping dominates over tidal effects and begins at fairly large cluster centric distances: 720 kpc (925 km s$^{-1}$) and 1.4 Mpc (1602 km s$^{-1}$). \label{tidal}}
\end{figure}

We now compute the relative strengths of the cluster's radial acceleration and the galaxy's centripetal acceleration. The stellar disk is detected out to $40\arcsec$ in radius by \cite{sun07}; this corresponds to a physical distance of 12.6 kpc, which we pick to be the radius of the disk. We estimate the galaxy's dynamical mass by using the K-band velocity-luminosity relation given in \cite{courteau07}. For its K-band luminosity, ESO137-001 should have a rotational velocity of 110 km s$^{-1}$, which coupled with its stellar disk size, corresponds to a dynamical mass of $3.5\times10^{10}\:M_\odot.$ In Figure \ref{tidal}, we show the ratio of the cluster's radial and galaxy's centripetal acceleration for the NFW and X-ray gas mass profile. Even at cluster-centric distance of 280 kpc, the minimum likely separation of the galaxy from the cluster center, we find that the cluster tidal field does not exceed the galaxy's centripetal acceleration. We also calculate the galaxy's tidal radius to be 15 kpc at this cluster-centric distance using the NFW profile for the cluster mass distribution. This is consistent with the value derived by \cite{sun07}. Therefore, cluster tidal forces are insufficient to cause any appreciable stripping of the stellar and gaseous components of the galaxy.

\subsubsection{Hydrodynamic Stripping}
\label{gasstripping}
We have shown that the molecular tail emanating from ESO 137-001 cannot result solely from galaxy interactions or from cluster tidal effects. We now explore the possibility that it arises from ram pressure stripping. The fact that the stellar disk appears relatively unperturbed signals that the tail must be generated by a mechanism that only affects gas. ICM ram-pressure stripping or viscous stripping are the most likely candidates. 
\par
To evaluate this possibility, the cluster-centric radius within which ram-pressure exceeds the galaxy's restoring force can be estimated using the \cite{fujita99} reformulation of the \cite{gunn72} criterion:
\begin{eqnarray}
\rho_{ICM}v_{gal}^2 & > & 2\pi G \Sigma_{star}\Sigma_{HI} \\
& > &2.11\times10^{-11}\:\textrm{dyne cm$^{-2}$}\left(\frac{v_{rot}}{220\:\textrm{km s$^{-1}$}}\right)^2\left(\frac{R}{10\:\textrm{kpc}}\right)^{-1} \nonumber \\
& & \times \left(\frac{\Sigma_{HI}}{8\times10^{20}m_H\:\textrm{cm$^{-2}$}}\right)
\end{eqnarray}
where $\rho_{ICM}$ is the density of the ICM, $v_{gal}$ is the velocity of the galaxy with respect to the ICM, $v_{rot}$ is the rotational velocity of the galaxy, $R$ is the radius of the disk, $\Sigma_{HI}$ is the surface density of the HI disk, and $m_H$ is the mass of the hydrogen atom. \cite{boehringer96}'s $\beta$-model fit \citep{cavaliere78} to their ROSAT X-ray data provides an estimate of $\rho_{ICM}.$ We use the $\beta$-model parameter values of $\rho_0 = 4.82\times10^{-27}\:\textrm{g cm$^{-3}$},$ $r_X = 184.5\:\textrm{kpc},$ and $\beta = 0.555$ to compute the ICM density as a function of cluster-centric radius. We also use the rotational velocity and radial extent of the galaxy given in Section \ref{clustertidalsec}. The HI surface density and the 3D velocity of this galaxy are unknown. \cite{broeils97} and \cite{karachentsev99} have shown that the HI surface density of spiral galaxies is fairly constant irrespective of the Hubble type or rotational velocity. \cite{karachentsev99} specifically looked at a nearby volume-limited sample of galaxies and found the HI surface density had a weak dependence on a galaxy's rotational velocity at less than the 1$\sigma$ level. The studies give slightly different mean surface densities. \cite{broeils97} compute an average surface density of $3.8\pm1.1\:\textrm{M$_\odot$ pc$^{-2}$}$ for their sample of spirals and irregulars.  From the HI surface density - rotational velocity relation shown in \cite{karachentsev99}, we estimate a surface density of $6.3\:\textrm{M$_\odot$ pc$^{-2}$}.$ For our calculations we were conservative and chose the higher surface density, as that increases the minimum pressure required for ram-pressure stripping. For the galaxy velocity, we picked the values that bound a galaxy's speed in a virialized cluster: $\sigma$ (925 km s$^{-1}$) and $\sqrt{3}\sigma$ (1602 km s$^{-1}$) where $\sigma$ is the velocity dispersion of the cluster. As an additional check we compute the Mach number of the galaxy using the X-ray data from \cite{sun06}. Assuming the galaxy is experiencing a shock as it travels through the ICM, we can predict the Mach number if we know the density and temperature of the head of the X-ray tail and the ICM. \cite{sun06} quote an average electron density and temperature for the head of the tail to be $1.8-3.5\times10^{-2} f^{-1/2}$ cm$^{-3}$ and 0.64 keV, respectively, where f is the filling factor. The ICM on the other hand has a temperature of $6.3$ keV and an electron density of $1.4\times10^{-3}$ cm$^{-3}$ at the location of the galaxy \citep{sun06}. If we assume a filling factor of 1, we obtain a Mach number that ranges from $1.1-1.6$. Using the ICM sound speed of approximately 800 km s$^{-1}$ that we calculate below, we obtain a galaxy velocity through the ICM of $880-1280$ km s$^{-1}.$ This matches closely with our assumed value, suggesting that the galaxy cannot be moving significantly faster than our predictions.
\par
For a galaxy traveling at a velocity of 925 km s$^{-1}$, ram-pressure stripping will become significant at cluster-centric radii less than 0.72 Mpc. However, if the galaxy is traveling at 1602 km s$^{-1}$ the radius increases to 1.4 Mpc. We show these results in Figure \ref{tidal} for the two velocity cases where we plot the effectiveness of ram-pressure in terms of the parameter $q_{ram}$ where $q_{ram} = \rho_{ICM}v_{gal}^2 / 2\pi G \Sigma_{star}\Sigma_{HI}$. When $q_{ram} > 1$, ram-pressure stripping becomes effective. Because ESO 137-001 has a very low line-of-sight velocity with respect to the cluster ($\sim 77\:\textrm{km s$^{-1}$}$), its true distance from the cluster center is not likely to be much larger than its projected distance. \cite{boselli06} state that infalling galaxies typically follow highly eccentric orbits. We can estimate the cluster-centric distance of the galaxy assuming a roughly linear orbit with the following equation:
\begin{equation}
r = \frac{R}{\cos(\sin^{-1}(v_{los}/v_{gal}))}
\end{equation}
where $r$ is the cluster-centric distance, $R$ is the projected distance from the cluster center, $v_{los}$ the line-of-sight velocity of the galaxy, and $v_{gal}$ the 3D velocity of the galaxy. The two galaxy velocity limits mentioned previously put the galaxy at a cluster centric distance of approximately 280 kpc. The galaxy is well within the minimum radius limits where ram-pressure should be a significant transformative process.
\par
We also consider the case where the galaxy is undergoing viscous stripping \citep{nulsen82}. This can be a significant stripping process with the same scaling properties as ram-pressure stripping if the gas flow is turbulent. For turbulent flow, the kinematics of the galaxy through a cluster should produce a Reynolds number $>30$. The Reynolds number, Re, is given in terms of the mean-free-path and sound speed in the ICM, $\lambda_{ICM}$ and $c_{ICM}$:
\begin{eqnarray}
Re &= &2.8\left(\frac{r_{gal}}{\lambda_{ICM}}\right)\left(\frac{v_{gal}}{c_{ICM}}\right) \\
c_{ICM} &=& \sqrt{\frac{kT}{m_H}} \\
\lambda_{ICM} &= &11\left(\frac{T_{ICM}}{10^8\:K}\right)^2\left(\frac{10^{-3}\:cm^{-3}}{\rho_{ICM}/m_H}\right)
\end{eqnarray}
where T is the temperature of the ICM. We assume an ICM temperature of 6.3 keV \citep{sun06}. We calculate Re for the galaxy located at 280 kpc from the cluster center for the two velocity limits. We obtain $30 < Re < 52$ for $925 < v_{gal} < 1602\:\textrm{km s$^{-1}$}$. We therefore expect that the gas flow across the wake of the galaxy is turbulent, as also shown in hydrodynamic calculations \citep[e.g.,][]{roediger08}. It is plausible that eddies in this part of the tail promote the star formation occurring there. We can also estimate the mass loss rate following the prescription in \cite{nulsen82}, which is $\dot{M} = \pi r^2 \rho v_{gal}$ where $r$ is the radius of the gas disk and $\rho$ is the density of the ICM. We set the radius of the gaseous disk to be the half-width of the 17.035 $\mu$m emission at the head of the warm H$_2$ tail, which is approximately 3 kpc. This is the best guess at the width of the current gas disk width. We obtain a mass loss rate that ranges from $0.7 < \dot{M} < 1.2$ M$_\odot$ yr$^{-1},$ which is a factor of 3 lower the value that we obtained from our warm molecular hydrogen observations (see Section \ref{massloss} for warm H$_2$ mass loss rate). This indicates that ram-pressure and not viscous stripping dominates, but viscous stripping may also contribute to the mass loss.

\subsubsection{Mass Loss Rate and Stripping Time Scale}
\label{massloss}
We calculate an instantaneous warm H$_2$ mass loss rate by dividing the measured warm H$_2$ mass (LL full tail region) within the 20 kpc long tail (seen in projection in the sky) by the time it would take the galaxy to traverse that distance. We adopt the galaxy velocity limits of 925 km s$^{-1} < v_{gal} <$ 1602 km s$^{-1}$ used in our ram-pressure calculation. The galaxy's velocity is taken to lie in the plane of the sky due to its low recessional velocity with respect to the cluster. It will take the galaxy approximately $1-2\times10^7$ years to traverse the length of the tail, which corresponds to a warm H$_2$ mass loss rate of $1.9 - 3.4$ M$_\odot$ yr$^{-1}.$ However, it is unclear what fraction of the total stripped gas the warm H$_2$ represents;  the mass loss rate presented is clearly a lower limit as evidenced by the H$\alpha$ and X-ray tails already detected. In order to make an estimate of the total gas mass loss rate, we need measurements of two important parameters: the total H$_2$ to warm H$_2$ mass ratio ($\gamma_{H_2}$) and the total gas to H$_2$ mass ratio ($\gamma_{gas}$). Accounting for these ratios yields a total mass loss rate of $(1.9 - 3.4)\gamma_{H_2}\gamma_{gas}$  M$_\odot$ yr$^{-1}.$ However, both of these $\gamma$ values are unknown for this galaxy's tail because no CO observations or HI measurements of the tail exist. X-ray and H$\alpha$ measurements do provide some constraints of the atomic hydrogen content of the tail, but the measured mass depends strongly on the filling factor, which is unknown.
\par
To estimate the true mass loss rate of the galaxy, we make a few simplifying assumptions to determine $\gamma$ values by using measurements made in other systems. $\gamma_{H_2}$ can be bound by cold-to-warm H$_2$ gas mass ratios determined by \cite{roussel07} for SINGS galaxies and by \cite{guillard10} for the shock front in Stefan's Quintet. In the SINGS sample, \cite{roussel07} on average measure a $\gamma_{H_2}$ of $\sim 10,$ while \cite{guillard10} measures a $\gamma_{H_2}$ of $\sim 2$ at the strongest portion of the galactic shock in Stefan's Quintet. This provides a plausible range of  $2 \lesssim \gamma_{H_2} \lesssim 10.$ It is likely that the gas in the shock front in Stefan's Quintet is in a much more extreme state than in our case, yielding a much smaller $\gamma_{H_2}$ ratio than what is likely for the tail of ESO 137-001. This possibility is strengthened by the rather large $\gamma_{H_2}$ of $\sim 10$ measured in NGC 4450, which is also thought to have a similar excitation mechanism as our H$_2$ tail \citep{roussel07}. H$_2$ is likely to be disassociated during the process of stripping, in which case the observed molecular hydrogen may form inside the tail after the shock-heated gas cools. In any case, we do not expect $\gamma_{H_2}$ to be a strong function of time.
\par
In comparison, it is plausible that $\gamma_{gas}$ has changed as the stripping has progressed. Initially, the large HI envelope of the galaxy would have presented a large cross section for stripping. There is even indirect evidence that the HI loss rate must have been high in the past. To illustrate, we estimate the initial gas mass of ESO 137-001 prior to its entering the cluster. We do so from the stellar mass of the galaxy, multiplied by the typical value of the gas to stellar mass ratio for a galaxy of the given K-band luminosity, M$_{gas}$ = 0.8 M$_*$, where M$_*$ is the stellar mass of the galaxy \citep{bell00}. The result is that the galaxy should have had an initial gas mass of $\sim$ 10$^{10}$ M$_\odot$, with an rms scatter of 0.2 dex \citep{bell00}. This mass is about an order of magnitude higher than the upper limit to the current HI content of ESO 137-001 \citep{sun07}, indicating that most of the gas has already been stripped. This is not surprising, since ram-pressure stripping is thought to become important at $3 - 5$ times the current cluster-centric distance of the galaxy. The time during which this stripping has occurred is at most $400-700$ Myr, the time to reach its current position at the range of estimates for the velocity of the galaxy from the radius where the ram-pressure criterion is met. 
\par
We can also estimate an upper limit for stripping the remaining gas. We combine the upper limit on the HI mass of 10$^9$ M$_\odot$ with an estimate of the initial H$_2$ mass as 0.1 M$_{gas}$ (the expected value for a Sc galaxy with $10^{10}$ M$_\odot$ of gas \citep{obr09}), or 10$^9$ M$_\odot$, to set an upper limit of $2 \times 10^9$ M$_\odot$ for the current gas content (HI plus H$_2$). From the loss rate of warm H$_2$ and the lower limit of $\sim$ 2 on $\gamma_{H2}$, there is a lower limit to the current mass loss of $3.8 - 6.8$ M$_\odot$ yr$^{-1}$ , even assuming all the lost gas is in molecular form (that is, ignoring the additional mass loss for gas in atomic form). We then find an upper limit of $300-500$ Myr to lose the remaining gas. This timescale is similar to the time it would take the galaxy to travel from its current position to the outskirts of the cluster core, which we take to be the core radius given by the X-ray gas distribution, for the range of galaxy velocities explored. This means the galaxy is currently roughly half way through the process of being completely stripped of its gas and will lose most of its gas before it finishes its first pass toward and then away from the cluster core.   
\par
This also yields an upper limit on the timescale for ram-pressure stripping of $\sim1$ Gyr, which is the sum of the time the galaxy has been experiencing ram-pressure stripping and the time it will take to lose all of its gas from now, for the range of galaxy velocities explored. There are several reasons why the true stripping timescale is much shorter: 1.) We assume no molecular gas has been stripped in our calculation of the time it would take for the galaxy to exhaust its gas from the present moment. 2.) The current measured HI mass used in our calculation is only an upper limit. 3.) We assume no atomic gas is currently being stripped, which does not explain the existence of both a X-ray and a H$\alpha$ tail. 4.) We do not account for the impact of star formation, which could be a significant consumer of gas. 5.) Even though the ram-pressure criterion may be met at large cluster-centric distances, the ICM wind may not be strong enough to strip the gas permanently from the galaxy's halo during the early stages of stripping. This is consistent with simulations of ram-pressure stripping, which indicate that significant gas stripping happens in a relatively quick timescale of a few hundred Myr for a galaxy of similar gas mass but 10 times more massive \citep{roediger05,kapferer09}. This may suggest that the warm H$_2$ may only represent a small fraction of the total gas stripped.

\subsection{Implications}
\label{implication}
The existence of H$_2$ in the ram-pressure stripped tail is puzzling because this is at odds with conventional wisdom that molecular gas is much more tightly bound than HI gas, making it far more difficult to strip than the less dense atomic gas. This is supported by the observation that cluster galaxies with strong HI deficiencies also have fairly normal amount of molecular gas content \citep{boselli06}. To determine how molecular hydrogen can exist outside the galaxy within the hot intracluster medium we carry out a simple calculation of the effect of ram-pressure on a typical molecular cloud. At the current location of ESO 137-001 where ram-pressure is significant, a spherical molecular cloud 10 pc in radius and $10^4$ M$_\odot$ in mass will only be moved by ram-pressure a few parsecs in a 1 Gyr. This means that ram-pressure cannot move whole clouds out of the galaxy. It is possible that molecular clouds may be ablated by the ICM wind and stripped gradually. However, the more likely explanation is that the high velocity dissociative shocks arising from the interaction between the ICM and the ISM of the galaxy dissociate the molecular hydrogen within the galaxy only to have it form on grain surfaces when the shocked gas cools \citep{guillard09}. It is unclear how the molecular gas is able to survive the harsh ICM environment. Our IRS observations have a resolution of $\sim 3$ kpc, insufficient to resolve any structure within the molecular hydrogen tail.  We do, however, see the strongest molecular hydrogen in the part of the tail that is closest to the galaxy, and its emission grows fainter the further away it is from the galaxy. This likely results from both the destruction of the molecular hydrogen by the ICM and the gradual loss of the kinetic energy from the ICM-ISM interaction. Given the velocity of the galaxy through the cluster, we can place a lower limit on the lifetime of the warm molecular hydrogen to be $10-20$ Myr. This number is likely to be larger if the molecular hydrogen exists in the H$\alpha$ and X-ray emitting parts of the tail, which extend to 40 kpc \citep{sun07} and 70 kpc \citep{sun06}, respectively. However, our data do not conclusively support shocks being the sole excitation mechanism as X-ray irradiation can also be a possible candidate.  Further ground-based observations characterizing the cold gas (CO and HI measurements) and hot H$_2$ gas (near-IR measurements of H$_2$ rovibrational lines) are required to fully understand this phenomenon. Explaining the existence of warm H$_2$ in the tail from ram-pressure stripping remains a big challenge as simulations are only beginning to understand the emission properties of the stripped gas tails, which do not include any molecular content (e.g. \cite{kapferer09,tonnesen10}).
\par
The star-forming regions seen outside of the plane of the galaxy, some of which are outside the tidal radius, are remarkable as a demonstration that ram-pressure stripping can be a source of intracluster stars. A few of these star-forming regions are likely associated with the H$_2$ tail indicating that in addition to the warm molecular gas there must also be cold molecular gas with sufficient density to support star formation. The turbulent flow of stripped multiphase ISM of the galaxy may allow for eddies to form in the wake where star formation can occur. This is consistent with ram-pressure simulations that do indeed show that significant star formation in high density clumps can occur in the wake of the galaxy \citep{kapferer09}. However, their simulations do not explicitly include molecular gas, so the picture painted by the simulation is incomplete and does not fully explain what is observed in ESO 137-001. It is entirely reasonable that the existence of significant molecular gas in the tail may enhance star formation within the tail, but a measurement of the cold gas fraction is necessary to verify this hypothesis.
\par
Next, we compare the morphology of the tail, the mass loss rate, and overall time scale for stripping with simulations to see if these properties are generally consistent. The collimated morphology of ESO 137-001's tail and its moderate mass loss rate indicate that it is in the intermediate stage of ram-pressure stripping, where the gas originally displaced from the gaseous disk of the galaxy is permanently removed from the halo of the galaxy \citep{roediger05}. This phase, which follows the short instantaneous stripping phase ($\sim 10$s Myr), lasts approximately a few 100 Myr in medium to strong ram-pressure. This is broadly consistent with our $< 1$ Gyr upper limit for the ram-pressure stripping timescale. However, these simulations do not explicitly include the effects of ram-pressure on molecular gas, so it is uncertain if the results of simulations can be directly compared. Moreover, it is not clear if H$_2$ stripping persists during the entire ram-pressure stripping phase. There is already some indication that the galaxy may be nearing the end of its ram-pressure stripping phase from its lower than expected HI gas mass and the small size of the H$\alpha$ disk within the galaxy.
\par
Our conclusion that the galaxy will lose all of its gas in a single pass through the core is also consistent with existing observational evidence. Assuming that a warm H$_2$ tail is found with either a X-ray or H$\alpha$ tail, a literature survey of galaxies with detected tails in H$\alpha$ \citep{gavazzi01,cortese06,yagi07,kenney08,yoshida08,sun07,sun10}, and/or X-ray \citep{wang04,sun05,sun06,sun10} reveals that these tails are rare. \cite{sun06} in their X-ray survey of 25 nearby hot clusters only discovered 2 galaxies with X-ray tails, although another case, ESO 137-002, has now been found serendipitously in Abell 3627 \citep{sun10}. However, this may not be indicative of the true distribution because the X-ray gas in the tail is much cooler than the ICM \citep{sun06} and may be impossible to discern from the hot ICM X-ray emission especially in the core regions where ICM X-ray emission is the highest. Moreover, the observed H$\alpha$ tails are low surface-brightness features that require relatively long integration times to detect them \citep{gavazzi01}. We also include the case where blue stellar trails were seen in galaxies within clusters \citep{cortese07}, which is similar to the star-forming trail seen in ESO 137-001. \cite{cortese07}, in their \emph{HST} survey of 13 different clusters at $z\sim 0.2$, only found two spiral galaxies with these types of trails. The rarity of these tails implies that the timescale for their existence is much shorter than the typical crossing time of a massive cluster, which is $\sim 5$ Gyr. This is broadly consistent with our timescale estimate, though these features may only be present when stripping is intense. If warm H$_2$ emission does not directly correspond to X-ray or H$\alpha$ emission or exists for a shorter time, it is entirely possible this phenomenon is even more rare.

\section{Conclusions}
We report on the detection of a warm H$_2$ tail in ESO 137-001, an infalling cluster spiral in Abell 3627. The main results of this study are:
\par
1. Through {\it Spitzer} IRS spectral mapping observations, we detect a warm molecular hydrogen tail (T $\sim 130-160$ K) extending at least 20 kpc from the galaxy ESO 137-001 in the cluster Abell 3627. The full extent of the tail is unknown and is limited by our slit coverage. The tail is co-aligned with both the X-ray \citep{sun06} and the H$\alpha$ \citep{sun07} tail. We measure a total warm H$_2$ mass of $4.11^{+0.43}_{-0.39}\times10^7$ M$_\odot$ within the H$_2$ tail. Approximately, 50\% of the warm H$_2$ tail mass is located outside the tidal radius of the galaxy and is permanently stripped.
\par
2. Multiple temperature components are present in the tail, and we fit a two-temperature-component model to the observed population distribution. We find that the hot component (T $= 472^{+77}_{-56}$ K) represents approximately 1\% of the gas mass of the tail in the region that we observed.
\par
3. We identify star-forming regions near the galaxy by identifying 8$\mu$m excesses (measured relative to 3.6$\mu m$), which coincide with most of the H$\alpha$ regions detected by \cite{sun07}. We also identify a large 8$\mu$m excess centered about the galaxy. Several of these star-forming regions exist outside the plane of the galaxy, some of which are beyond the tidal radius. Therefore, ram-pressure stripping can also be a source of intracluster stars.
\par
4. The molecular hydrogen tail extends farther than the extent of the star-forming regions. A comparison of the H$_2$ line luminosity with 24 $\mu m$ emission strongly suggests that star formation is not the main excitation mechanism for the warm H$_2$ tail.
\par
5. The infrared spectrum of ESO 137-001 itself matches expectations for a normal starforming galaxy. We estimate a star-forming rate at the nucleus of $0.7$ M$_\odot$ yr$^{-1}.$
\par
6. Assuming a velocity range of $925 (\sigma) < v_{gal} < 1602 (\sqrt{3}\sigma)$ km s$^{-1}$ for the galaxy, we obtain a warm molecular hydrogen mass loss rate of $1.9-3.4$ M$_\odot$ yr$^{-1}.$ We place a strong upper limit on the time it would take to strip the remaining gas in the galaxy of $300-500$ Myr and one on the total ram-pressure stripping time scale of 1 Gyr. This means that the galaxy will have lost most of its original gas content after a single pass through the cluster core, and that ram-pressure operates on timescales much shorter than the crossing time in a cluster. These results are consistent with the hypothesis that ram-pressure stripping can transform galaxies quickly when they fall into the cores of massive clusters.

\acknowledgements

We would like to thank Ming Sun for providing us with his H$\alpha$ data and giving several useful suggestions for the paper. We thank Lisa Storrie-Lombardi for expediting obtaining our follow-up SL spectra. We also thank the anonymous referee whose comments were very useful in strengthening the paper. This work was supported by contract 1255094 from Caltech/JPL to the University of Arizona. This research has made use of the NASA/IPAC Extragalactic Database (NED) which is operated by the Jet Propulsion Laboratory, California Institute of Technology, under contract with the National Aeronautics and Space Administration.

\newpage

\newpage
\clearpage

\input{tab1.tex}

\input{tab2.tex}

\input{tab3.tex}

\input{tab4.tex}

\input{tab5.tex}

\end{document}

%% file: tab1.tex
\begin{deluxetable}{lcccc}
\tablewidth{0pt}
\small
\tablecolumns{5}
\tablecaption{Measured H$_2$ Rotational Line Fluxes for ESO 137-001 Tail \label{wavelengths}}
\tablehead{
\colhead{Spectral Feature} & 
\colhead{$\lambda$} &
\colhead{Tail (SL/LL) Flux} &
\colhead{All Tail (LL-only) Flux} &
\colhead{Far Tail (LL-only) Flux} \\
&  ($\mu m$) & ($10^{-17}$ W m$^{-2}$) & ($10^{-17}$ W m$^{-2}$) & ($10^{-17}$ W m$^{-2}$)
}
\startdata
0-0 S(7) & 5.511 & $1.03\pm$0.26 & --- & ---\\
0-0 S(6) & 6.109 &  $<0.74$\tablenotemark{*} & --- & ---\\
0-0 S(5) & 6.909 &  $<1.04$\tablenotemark{*} & --- & ---\\
0-0 S(4) & 8.026 &  $<0.59$\tablenotemark{*} & --- & ---\\
0-0 S(3) & 9.665 &  $1.20\pm0.11$ & --- & ---\\
0-0 S(2) & 12.279 & $0.42\pm0.08$ & --- & ---\\
0-0 S(1) & 17.035 & $0.92\pm0.01$ & $2.10\pm0.08$ & $0.80\pm0.09$\\
0-0 S(0) & 28.221 & $0.23\pm0.01$ & $0.47\pm0.03$ & $0.21\pm0.02$
\enddata
\tablenotetext{*}{3$\sigma$ detection limit.}
\end{deluxetable}

%% file: tab2.tex
\begin{deluxetable}{lcccc}
\tablewidth{0pt}
\small
\tablecolumns{5}
\tablecaption{Fine Structure Line Fluxes for ESO 137-001 Tail \label{finestructure}}
\tablehead{
\colhead{Spectral Feature} & 
\colhead{$\lambda$} &
\colhead{Tail (SL/LL) Flux} &
\colhead{Full Tail (LL-only) Flux} &
\colhead{Far Tail (LL-only) Flux} \\
&  ($\mu m$) & ($10^{-17}$ W m$^{-2}$) & ($10^{-17}$ W m$^{-2}$) & ($10^{-17}$ W m$^{-2}$)
}
\startdata
{[}NeII] & 12.8 & $1.17\pm0.11$ & --- & ---\\
{[}NeIII] & 15.6 & $0.25\pm0.01$  & $0.17\pm0.08$ & $< 0.21$\\
{[}SIII] & 18.7  & $0.30\pm0.01$ & $0.34\pm0.08$ & $0.17\pm0.04$\\
{[}OIV],[FeII]\tablenotemark{\dag} & 25.9, 26.0 & $0.14\pm0.01$  & $< 0.10\tablenotemark{*}$ & ---\\
{[}SIII] & 33.5 & $0.87\pm0.02$ & $0.90\pm0.10$ & $0.18\pm0.06$ \\
{[}SiII] & 34.8 & $1.16\pm0.02$ & $1.35\pm0.11$ & $0.57\pm0.08$
\enddata
\tablenotetext{\dag}{Possible blended feature.}
\tablenotetext{*}{3$\sigma$ detection limit.}
\end{deluxetable}

%% file: tab3.tex
\begin{deluxetable}{cccccc}
\tablewidth{0pt}
\small
\centering
\tablecolumns{5}
\tablecaption{Measured H$_2$ Gas Masses for ESO 137-001  \label{gasmass}}
\tablehead{
\colhead{Fit} & 
\colhead{Component} &
\colhead{T$_{\rm ex}$ (K)} &
\colhead{$N_{\rm tot}$ (cm$^{-2}$)} &
\colhead{$\Sigma$ (M$_\odot$ pc$^{-2}$)} &
\colhead{H$_2$ Mass (M$_\odot$)}
}
\startdata
Full Tail (LL-only) & Warm &$157\pm3$ & $2.56^{+0.27}_{-0.24}\times10^{19}$ & $0.41^{+0.04}_{-0.04}$ & $4.11^{+0.43}_{-0.39}\times10^7$ \\
\tableline
Far Tail (LL-only) & Warm &$149\pm6$ & $2.90^{+0.51}_{-0.43}\times10^{19}$ & $0.46^{+0.08}_{-0.07}$& $2.15^{+0.38}_{-0.32}\times10^7$\\
\tableline
Tail (SL/LL) & Warm & $125^{+10}_{-19}$ & $5.9^{+4.2}_{-1.1}\times10^{19}$ & $0.94^{+0.67}_{-0.18}$ & $3.6^{+2.6}_{-0.7}\times10^7$ \\
& Hot & $472^{+77}_{-56}$ & $5.9^{+4.4}_{-2.6}\times10^{17}$ & $9.4^{+7.0}_{-4.1}\times10^{-3}$ & $3.6^{+2.7}_{-1.6}\times10^5$ \\
\enddata
\end{deluxetable}

%% file: tab4.tex
\begin{deluxetable}{lcccc}
\tablewidth{0pt}
\small
\tablecolumns{5}
\tablecaption{Dusty 8$\mu m$ Source List\label{8umexcess}}
\tablehead{
\colhead{\#} & 
\colhead{RA} &
\colhead{DEC} &
\colhead{Distance ($\arcsec$)} &
\colhead{H$\alpha$ counterpart}}
\startdata
1 & 16:13:27.19 & -60:45:50.44 &   0.8 & Y \\
2 & 16:13:24.98 & -60:45:43.51 &  18.3 & Y \\
3 & 16:13:26.48 & -60:45:32.85 &  18.6 & Y \\
4 & 16:13:25.45 & -60:45:34.79 &  20.7 & Y \\
5 & 16:13:23.93 & -60:45:53.60 &  24.9 & Y$^{\tablenotemark{*}}$ \\
6 & 16:13:24.54 & -60:46:05.92 &  25.4 & Y \\
7 & 16:13:24.13 & -60:45:33.46 &  28.8 & Y \\
8 & 16:13:23.66 & -60:45:23.70 &  37.8 & Y \\
9 & 16:13:25.78 & -60:45:10.21 &  41.8 & Y \\
10 & 16:13:29.82 & -60:46:30.42 &  44.0 & N \\
11 & 16:13:22.83 & -60:45:20.55 &  44.4 & Y \\
12 & 16:13:23.03 & -60:44:49.45 &  68.6 & -\\
13 & 16:13:35.89 & -60:45:23.07 &  68.6 & Y \\
14 & 16:13:33.74 & -60:44:49.60 &  77.0 & N\\
15 & 16:13:32.01 & -60:47:00.30 &  77.9 & N\\
16 & 16:13:36.16 & -60:44:53.67 &  86.2 & -\\
17 & 16:13:32.05 & -60:44:27.37 &  90.1 & N\\
18 & 16:13:30.07 & -60:47:25.82 &  97.5 & -\\
19 & 16:13:13.65 & -60:46:04.68 & 101.0 & N\\
20 & 16:13:39.49 & -60:44:47.56 & 109.3 & N\\
21 & 16:13:42.58 & -60:46:28.59 & 118.3 & -\\

\enddata
\tablenotetext{*}{This single source detection may actually be two separate sources.}
\end{deluxetable}

%% file: tab5.tex
\begin{deluxetable}{lcc}
\tablewidth{0pt}
\small
\tablecolumns{4}
\tablecaption{Aromatic Feature Fluxes \label{aromatic}}
\tablehead{
\colhead{Aromatic Feature} & 
\colhead{Nuclear Flux} &
\colhead{Tail (SL/LL Region) Flux} \\
  & ($10^{-17}$ W m$^{-2}$) & ($10^{-17}$ W m$^{-2}$)
}
\startdata
6.2 $\mu$m & $66.1\pm1.8$ & $2.21\pm0.72$ \\
7.7 $\mu$m Complex & $217\pm7$ & $21.0\pm2.1$ \\
11.3 $\mu$m Complex & $57.8\pm0.9$ & $3.43\pm0.29$ \\
12.6 $\mu$m Complex & $32.9\pm0.9$ & $0.88\pm0.35$ \\
17.7 $\mu$m Complex & $28.1\pm0.3$ & $1.76\pm0.07$
\enddata

\end{deluxetable}